\documentclass[trackchanges]{aastex701}
\usepackage{amsmath}
\usepackage{graphicx}
\usepackage{url}
\usepackage{multirow}
\usepackage{ragged2e}
\usepackage{booktabs}
\usepackage{subfigure}
\usepackage{longtable} 
\usepackage{CJK}
\usepackage{natbib}

\begin{document}
\defcitealias{2022ApJ...925...70Z}{Z22}
\title{Morphology of Optical Changing-Look AGN-host Galaxies: Evidence for an Important Role of Mergers}
\begin{CJK*}{UTF8}{gbsn} 
\author[0009-0001-3432-480X,gname=Jie,sname=Tian]{Jie Tian (田杰)}
\affiliation{Yunnan Observatories, Chinese Academy of Sciences, Kunming 650216, People's Republic of China}
\affiliation{University of Chinese Academy of Sciences, Beijing 100049, People's Republic of China}
\email[show]{tianjie@ynao.ac.cn}
\author[0000-0002-9128-818X,gname=Yinghe,sname=Zhao]{Yinghe Zhao (赵应和)}
\affiliation{State Key Laboratory of Radio Astronomy and Technology, Yunnan Observatories, Chinese Academy of Sciences, Kunming 650216, People's Republic of China}
\email[show]{zhaoyinghe@ynao.ac.cn}
\author[gname=Jin-Ming,sname=Bai]{Jin-Ming Bai}
\affiliation{Yunnan Observatories, Chinese Academy of Sciences, Kunming 650216, People's Republic of China}
\email[show]{baijinming@ynao.ac.cn}
\author[0000-0001-9457-0589,gname=Wei-Jian,sname=Guo]{Wei-Jian Guo}
\affiliation{National Astronomical Observatories, Chinese Academy of Sciences, 20A Datun Road, Chaoyang District, Beijing 100101, People’s Republic of China}
\email[show]{guowj@bao.ac.cn}
\correspondingauthor{Yinghe Zhao}
\email{zhaoyinghe@ynao.ac.cn}

\begin{abstract}

Optical changing-look active galactic nuclei (CL-AGNs) are characterized by the (dis)appearance of broad emission lines on unexpectedly short timescales. However, the underlying mechanisms and their potential connection to host-galaxy properties are still unclear. In this work, we present an analysis of the morphology for 63 low-redshift CL-AGNs ($z<0.15$) selected from the largest CL-AGN catalog (\citealt{2025ApJS..278...28G}) to date, using images from DESI DR10 and employing both non-parametric methods and visual inspection. We find that CL-AGN hosts exhibit a concentration like late-type spirals, asymmetry like early-type spirals, and smoothness like ellipticals. This is confirmed by their $Gini-M_{20}$ coefficients, suggesting weak/modest disturbances. Based upon our visual inspection, we further identify that 18 (29\%) out of 63 sources are mergers, among which $\sim$56\% (10/18) show shell features. Compared to different non-CL-AGN samples, CL-AGN hosts have a higher ($\sim$2$\times$) possibility of being merging systems. Our results indicate that mergers/interactions may play an important role in driving the changing-look behavior.

\end{abstract}

\section{INTRODUCTION} \label{sec:sec1}

Supermassive black holes (SMBHs) are typically found at the centers of active galaxies, where they release large amounts of energy by accreting surrounding material \citep{1995ARA&A..33..581K,1998AJ....115.2285M,2013ARA&A..51..511K}. Active galactic nuclei (AGNs) and quasars are the extreme cases of energy release from these black holes. They show significant radiation fluctuations across a broad range of the electromagnetic spectrum, from gamma rays to radio bands \citep{1984ARA&A..22..471R,2003MNRAS.345.1271V,2008ARA&A..46..475H,2010ApJ...716...30A,2017A&ARv..25....2P}.

In addition to the observed luminosity fluctuations, some AGNs show behavior distinct from the typical AGN emission spectrum, particularly concerning their broad emission lines (BELs). This behavior is referred to as the changing-look (CL) phenomenon. In these objects, the BELs may suddenly appear (turn-on) or disappear (turn-off) on timescales of months to years, which corresponds to spectral transitions between Type 1, intermediate, and Type 2 AGNs \citep{2015ApJ...800..144L,2016MNRAS.457..389M,2018ApJ...862..109Y,2023NatAs...7.1282R,2024ApJS..270...26G,2025ApJ...994..203S}. Such drastic changes challenge the traditional scenario of the unification model \citep{1993ARA&A..31..473A,1995PASP..107..803U,2015ARA&A..53..365N} and the widely accepted viscous heating timescale of hundreds of years for a steady accretion disk \citep{2010A&A...509A.106S,2016MNRAS.455.1691R,2017MNRAS.470.4112G,2018MNRAS.480.3898N}. As their spectra exhibit clear starlight features in the dim state, CL-AGNs also provide an excellent opportunity to explore AGN-host galaxy relationships, such as coevolution and feedback \citep{2017ApJ...835..144G,2019ApJ...883...31F}.

Recent studies have shown that CL-AGNs typically display low average Eddington ratios (approximately  $\lambda _{Edd}  = 0.01$), a crucial factor influencing CL-AGN events \citep{2019ApJ...874....8M,2022ApJ...933..180G,2024ApJ...966..128W,2025ApJS..278...28G}. In a low-Eddington-ratio ($<0.01$) AGN with an unstable inner disk, a temporary gas supply can drive an accretion rate change beyond the disk instability prediction and reionize the broad line region (BLR; \citealt{1998ApJ...504..671K}). Furthermore, \cite{2023ApJ...956..137W} suggest that CL-AGNs may represent a specific evolutionary phase of AGNs, transitioning from an abundant gas supply (``feast'') to a stage of fuel scarcity (``famine''). Although several causes (e.g., see the review by \citealt{2023NatAs...7.1282R}) are proposed to explain the changes in CL-AGNs, the underlying mechanisms and especially their potential link to host-galaxy properties are not well understood yet. 

The investigation of CL-AGN host galaxies, such as star formation histories (SFHs), stellar populations (SSPs), star-forming activities (SF), etc., can provide important clues to the mechanisms responsible for the CL phenomenon in AGNs. Despite a growing number of works 
(e.g., \citealt{2019ApJ...876...75C,2020MNRAS.498.3985Y,2021ApJ...907L..21D,2021ApJ...915...63L,2022ApJ...926..184J,2023ApJ...956..137W,2025ApJ...980...91Y,2025ApJ...989..101V}) that have studied the host galaxies of CL-AGNs, it is still unclear whether the properties on galactic scales can affect the small-scale accretion processes that drive the large variability of CL-AGNs. This might be caused by different challenges and uncertainties (e.g., AGN contamination and sample size) in different works. Furthermore, the vast majority of the aforementioned works focus on the stellar populations and star-forming properties of the hosts, while little attention has been paid to study the morphology of CL-AGN host galaxies. 

The morphology of a galaxy may reveal the assembly history, interaction with the environment and internal perturbation, and thus can give critical insights into its evolution. The only detailed analysis of CL-AGN hosts was presented by \cite{2019ApJ...876...75C}, who investigated the host morphology of four turn-off quasars and found that three of them are recent or ongoing mergers. However, the result may suffer from a large uncertainty of small sample sizes. 

In this paper, we present our investigation on the morphological properties for a homogeneous sample of 63 objects extracted from the largest spectroscopically identified CL-AGN catalog to date (\citealt{2025ApJS..278...28G}). We take advantage of non-parametric methods to quantitatively measure galaxy structure, and perform visual inspection to identify merging/interacting features. By further comparing CL-AGN hosts with those of non-CL-AGNs (NCL-AGNs; defined as AGNs left after removing our CL-AGNs from the normal AGNs in \citet[][hereafter Z22]{2022ApJ...925...70Z}), we also aim to uncover the potential connection between the CL-AGN phenomena and their host-galaxy properties, thereby shedding new light into the driver of CL events.

The structure of the paper is organized as follows. Section 2 outlines the sample selection and imaging data used in this study, providing a detailed description of the procedures involved in deriving non-parametric coefficients using \texttt{statmorph}\footnote{\url{https://statmorph.readthedocs.io/en/latest/overview.html}} \citep{2019MNRAS.483.4140R}. We also conduct a visual classification of CL-AGN host galaxy morphology. In Section 3, we analyze the sample data, compare our findings with those of NCL-AGN hosts, and examine the similarities and differences between the ``turn-on'' and ``turn-off'' stages of CL-AGN host galaxies. Finally, Section 4 presents a summary and conclusion. Where required, we adopt a Hubble constant of $H_0 = 73$~km~s$^{-1}$~Mpc$^{-1}$ as used in our previous work.

\section{SAMPLE, DATA AND METHOD} \label{sec:sec2}
\subsection{Sample} \label{sec:sec2.1}

\cite{2025ApJS..278...28G} identifies CL-AGNs from the Dark Energy Spectroscopic Instrument First Data Release (DESI DR1; \citealt{2026AJ....171..285D}) and Sloan Digital Sky Survey Data Release 16 (SDSS DR16; \citealt{2020ApJS..249....3A}) at $z<0.9$, confirming them through spectral flux calibration assessment using [O III]-based calibration, pseudophotometry (applying CRTS/PS1/ZTF filters to SDSS/DESI spectra), and visual inspection. Their catalog of CL-AGNs contains 561 objects, providing a wealth of samples for our morphological study.

In this work, we select a subsample of CL-AGNs with redshifts $z<0.15$ from the \cite{2025ApJS..278...28G} catalog. The cutoff of $z$ here is implemented to mitigate the challenges associated with subsequent non-parametric measurements and visual classifications for sources located at higher redshifts, as \cite{2018MNRAS.473.2701D} found that we can make reliable morphological measurements of galaxies for $z<0.2$ in images with a full width at half-maximum (FWHM) of the point spread function (PSF) about 1\arcsec\ (see also \citetalias{2022ApJ...925...70Z}). The resulting sample comprises 63 CL-AGNs.

\subsection{Images and Preprocessing} \label{sec:sec2.2}
We acquire images from the Dark Energy Spectroscopic Instrument (DESI) Data Release 10 (DR10)\footnote{\url{https://www.legacysurvey.org/dr10/description/}}, which includes the Dark Energy Camera Legacy Survey (DECaLS), Beijing-Arizona Sky Survey (BASS), and Mayall \textit{z}-band Legacy Survey (MzLS). The DR10 images come from the DECaLS \textit{grz} bands observations. The optical depth is defined as the median $5\sigma$ point-source depth (AB magnitudes), based on the formal errors in the Tractor catalogs for point sources. The predicted depths for 2 observations at 1.5\arcsec\ seeing are \textit{g} = 24.0, \textit{r} = 23.4, and \textit{z} = 22.5 mag \citep{2019AJ....157..168D}, approximately 2 mag deeper than SDSS imaging, which enables the identification of low surface brightness features. All 63 sample images are available in DESI DR10, and we use the right ascension and declination from the \cite{2025ApJS..278...28G} catalog as the center of the images. We choose the \textit{grz} band data to measure the morphological parameters, using cutouts of 512 pixels and a pixel scale of 0.262\arcsec\ across all three bands. DESI DR10 also provides the PSF for individual exposures, computed independently for each CCD using \texttt{PSFEx} \citep{2011ASPC..442..435B}. More details about the PSF are available on the DESI website\footnote{\url{https://www.legacysurvey.org/dr10/psf/}}. The FWHM of the PSF is approximately 1.18\arcsec\ in the \textit{r} data \citep{2019AJ....157..168D}. 

\begin{figure}[t]
    \centering
    \subfigure[]{
        \includegraphics[width=0.23\textwidth]{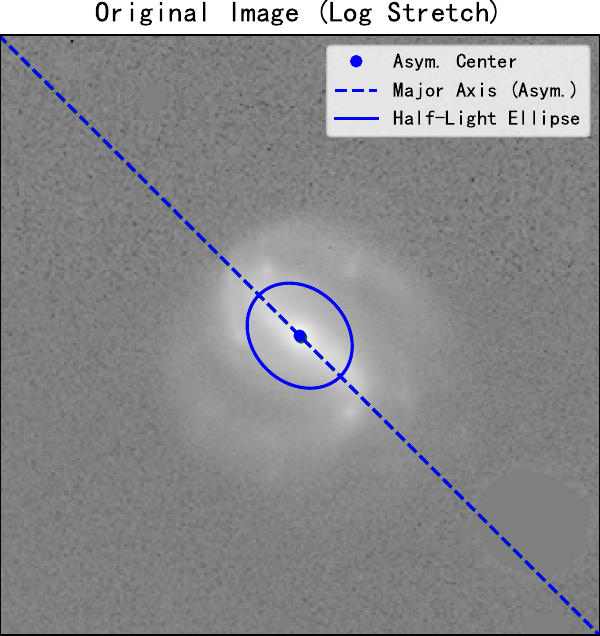}
    }
    \subfigure[]{
        \includegraphics[width=0.23\textwidth]{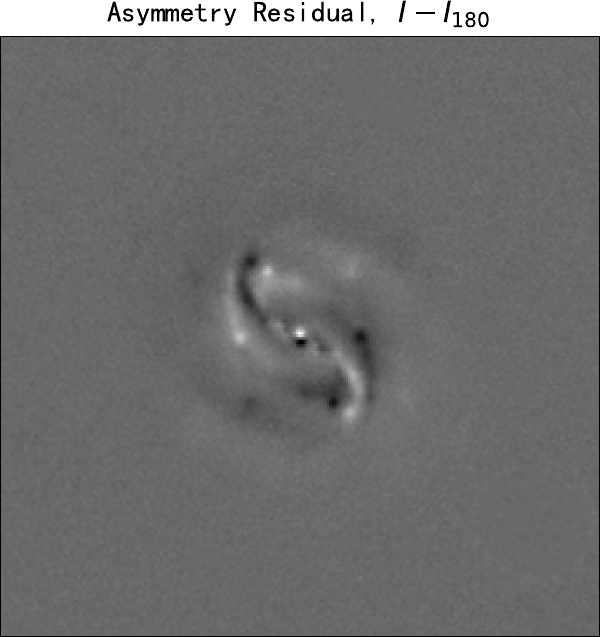}
    }
    \subfigure[]{
        \includegraphics[width=0.23\textwidth]{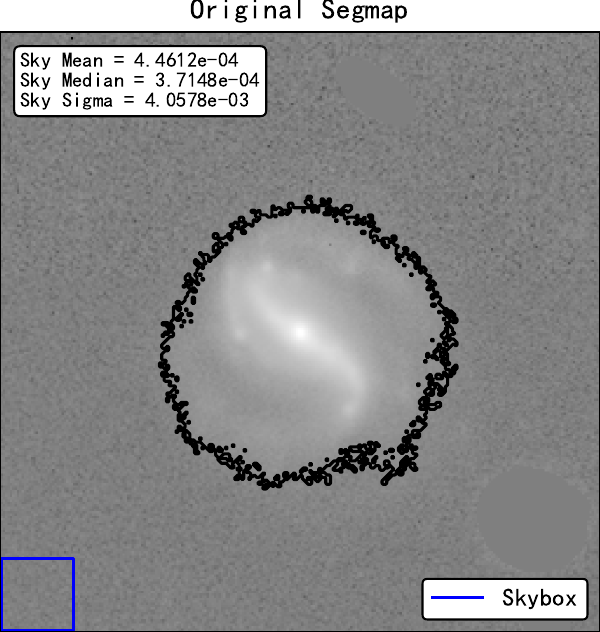}
    }
    \subfigure[]{
        \includegraphics[width=0.23\textwidth]{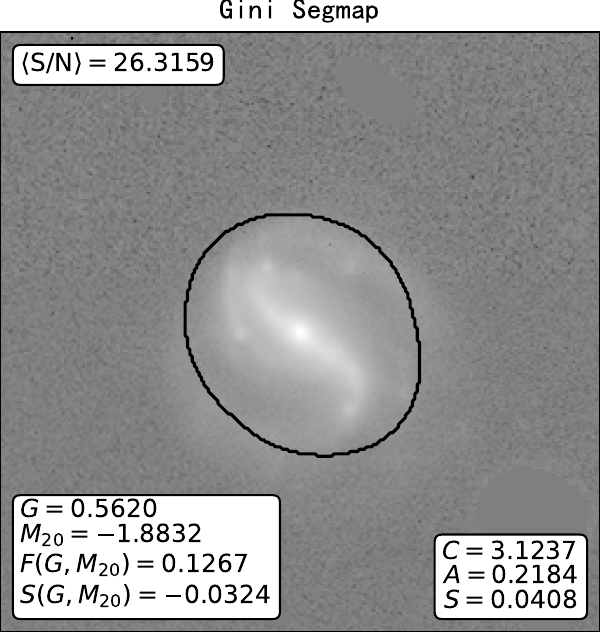}
    }
    \caption{Output plot of object ID55. (a) Original Image shows a cutout of the original galaxy image. The blue dot marks the galaxy center (defined as the point that minimizes asymmetry), while the blue dashed and solid lines indicate the orientation and extent of the elliptical aperture that encloses half of the total flux. (b) Asymmetry Residual shows the difference between the original image and its $180^{\circ}$ rotated version. (c) Original Segmap shows the segmentation map. Black contour outlines the galaxy, and the blue square marks the skybox for background. (d) \texttt{Gini\_Segmap} plot displays the $CAS$ and Gini-$M_{20}$ values. The black contour indicates the pixels used in their calculation, while the black solid line shows the Gini segmentation at $\eta = 0.2$. Increasing $\eta$ in the \texttt{Gini\_Segmentation} ignores substructures (e.g., spiral arms) but only slightly affects the measured values \citep{2022ApJ...925..157Z}.}
    \label{fig:fig5}
\end{figure}

In the image preprocessing, we employ \texttt{SEXtractor} \citep{1996A&AS..117..393B} and \texttt{Photutils} \citep{2020zndo...4049061B} to isolate the target from the background and generate the segmentation map, respectively. These steps follow the methods outlined in \cite{2025ApJ...986...66T}. First, we establish a detection threshold for objects $2 \sigma$ above the background to effectively eliminate contamination. Next, we designate a clean zone spanning 240 to 270 pixels to ensure that the target remains unaffected. After conducting a visual inspection, we mask these sources with a suitable radius to guarantee the complete absence of contamination. The mask radius ranges from 3 to 6 times the object's semi-major and semi-minor axes. Finally, we set the threshold to either 2 or $3\sigma$ above the background and $npixels = 10$ to generate the segmentation map. Nearby contamination may still affect some targets, requiring threshold adjustments for both the mask and the segmentation map. Figure \ref{fig:fig5} presents example output plots of the measurement.

After conducting the morphological measurements, we find no significant distinctions among the majority of parameters among the three bands (also see \citealt{2025ApJ...986...66T}). A Kolmogorov-Smirnov (K-S) test on \textit{g}-, \textit{r}-, and \textit{z}-band distributions further confirms this: except for the asymmetry ($A$) and $M_{20}$ between \textit{g}- and \textit{z}-band, and $A$ between \textit{r}- and \textit{z}-band (all having \textit{p}-values $< 0.05$), all the remaining parameters have \textit{p}-values $> 0.05$. Therefore, we only present the results from the \textit{r}-band data in the following analysis.

\subsection{Non-parametric Measurements} \label{sec:sec2.3}

Non-parametric morphological coefficients are essential tools for understanding the light distribution within galaxies. These coefficients not only define the spatial distribution of light but also aid in the exploration of evolutionary and structural characteristics of different galaxy types. Here, we exploit the Python package \texttt{statmorph} \citep{2019MNRAS.483.4140R} to measure several parameters widely used to quantify the structure of a galaxy, including concentration, asymmetry, and smoothness (i.e., the CAS system; see, e.g., \citealt{1995ApJ...451L...1S,1996MNRAS.279L..47A,2000AJ....119.2645B,2000ApJ...529..886C,2003ApJS..147....1C}), $M_{20}$ (\citealt{2004AJ....128..163L}), and the Gini coefficient (\citealt{2004AJ....128..163L}).

The concentration parameter (\textit{C}) is defined as the ratio between the radii containing 80\% ($r_{80}$) and 20\% ($r_{20}$)of the Petrosian \citep{1976ApJ...209L...1P} luminosity, the asymmetry parameter (\textit{A}) is obtained by subtracting the image rotated by 180 degrees from the original image, and the smoothness parameter (\textit{S}; also referred to as ``Clumpiness'') represents the clumpy structure within the galaxy. For the Gini coefficient (\textit{G}), it is defined by the Lorentz curve of the galaxy's light distribution, and a larger \textit{G} indicates a more concentrated profile. The $M_{20}$ statistic also reflects the concentration of a galaxy, but it is more sensitive than \textit{C} to merging signatures, such as multiple nuclei and bright tidal tails (\citealt{2003ApJS..147....1C}).

To ensure the reliability of all data, we establish the following selection criteria, which are consistent with those outlined in \cite{2025ApJ...986...66T}:
\begin{enumerate}
    \item \texttt{flag} == 0 and \texttt{flag\_sersic} == 0. (\texttt{statmorph} provides these two criteria to mark good data.)
    \item The Mask part does not take up too much area of the whole galaxy. (It ensures all the contamination can be removed cleanly, or does not affect the measurement of the target.)
    \item $r_{20}$ is bigger than half of the FWHM of the PSF. 
    \item sn\_per\_pixel is bigger than 2.5. 
\end{enumerate}

After applying these selection criteria, there remain 56 galaxies for our further quantitative analysis. The detailed measurements are provided in Table \ref{tab:tab3} in Appendix \ref{sec:A1}, and the mean and median values of the parameters are presented in Table \ref{tab:tab1}. Among the seven sources failed to measure their structural parameters, three galaxies are due to the influence of nearby contamination (criterion 2), another three samples are caused by their relatively low $r_{20}$ values (criterion 3), and the remaining one object shows \texttt{flag} == 1, indicating a problem of the segmentation map.

\begin{table}[t]
\movetableright=-35mm
\renewcommand{\arraystretch}{2} 
\caption{Mean and median values of structural parameters for different (sub-)samples}
\label{tab:tab1}
\resizebox{1\columnwidth}{!}{
\begin{tabular}{cccccccccccccc}\\\hline \hline
\multirow{2}{*}{Sample}&\multirow{2}{*}{Method}&\multicolumn{2}{c}{Number}&\multicolumn{2}{c}{Concentration}&\multicolumn{2}{c}{Asymmetry}&\multicolumn{2}{c}{Smoothness}&\multicolumn{2}{c}{Gini}& \multicolumn{2}{c}{$M_{20}$}\\\cline{3-14}
                         &&number&fraction$\pm \sigma$&mean($\sigma$)&median($\sigma$)&mean$(\sigma)$&median($\sigma$)&mean($\sigma$)&median($\sigma$)& mean($\sigma$)&median($\sigma$)&mean($\sigma$)&median($\sigma$)\\\hline
\multirow{2}{*}{total}   &statmorph&56&100\%&3.3(0.4) & 3.2(0.4)&0.06(0.05) & 0.04(0.03)&0.01(0.02) & 0.01(0.02)&0.54(0.04) &                 0.54(0.04)&-2.0(0.2) & -2.0(0.1)\\
&visual&63&100\%&\nodata & \nodata&\nodata& \nodata&\nodata& \nodata&\nodata &\nodata&\nodata & \nodata \\\hline
\multirow{2}{*}{merger}  &statmorph&15&$26.8\pm5.9\%$&3.3(0.4) & 3.3(0.5)&0.06(0.03) & 0.05(0.03)&0.01(0.01) & 0.01(0.01)&0.55(0.04) & 0.55(0.04)&-2.0(0.2) & -1.9(0.2) \\
 &visual&18&$28.6\pm5.6\%$&\nodata & \nodata&\nodata& \nodata&\nodata& \nodata&\nodata &\nodata&\nodata & \nodata  \\\hline
\multirow{2}{*}{non-merger} &statmorph&41&$73.2\pm5.9\%$  &3.2(0.4) & 3.2(0.4)&0.06(0.05) & 0.04(0.04)&0.02(0.02) & 0.01(0.02)&0.54(0.04) & 0.54(0.04)&-1.9(0.2) & -2.0(0.1)\\
&visual&45&$71.4\pm5.6\%$&\nodata & \nodata&\nodata& \nodata&\nodata& \nodata&\nodata &\nodata&\nodata & \nodata\\\hline
\multirow{2}{*}{turn-on} &statmorph&9&$16.1\pm4.9\%$  &3.6(0.3) & 3.6(0.4)&0.07(0.05) & 0.05(0.04)&0.01(0.01) & 0.01(0.01)
                         &0.57(0.03) & 0.55(0.03)&-2.1(0.1) & -2.1(0.1)\\
                         &visual&10&$15.9\pm4.6\%$&\nodata & \nodata&\nodata& \nodata&\nodata& \nodata&\nodata &\nodata&\nodata & \nodata\\ \hline
\multirow{2}{*}{turn-off} &statmorph&47&$83.9\pm4.9\%$  &3.2(0.4) & 3.2(0.3)&0.06(0.05) & 0.04(0.04)&0.01(0.02) & 0.01(0.02)
                          &0.54(0.04) & 0.54(0.04)&-1.9(0.2) & -1.9(0.1)\\
                          &visual&53&$84.1\pm4.6\%$&\nodata & \nodata&\nodata& \nodata&\nodata& \nodata&\nodata &\nodata&\nodata & \nodata\\\hline
\end{tabular}
}
\tablecomments{For each (sub-)sample, the first row lists the results for the sample of 56 sources with successful measurements from \texttt{statmorph}, and the second row shows the whole sample of 63 CL-AGNs with our visual classification. For the median, 1$\sigma$ dispersion is estimated using $1.4826\times \mathrm{MAD}$, where MAD is the median absolute deviation of the sample. The $1\sigma$ uncertainty for each fraction is calculated with the method given in \cite{Cameron_2011}.}
\end{table}

\subsection{Morphological Classification via Visual Inspection} \label{sec:sec2.4}

We also employ visual classification to distinguish between mergers and non-merging galaxies among all 63 CL-AGN host galaxies. To enhance the detail in the images for better classification, we create an ``unsharp masked'' version of all the images. Unsharp masking \citep{1977AASPB..16...10M} is a linear image processing technique that enhances sharp details by adding the difference between the original image and its blurred version. This method has been used to detect low surface brightness features like shells and other tidal features in nearby galaxies (e.g., \citealt{1983ApJ...274..534M}). The amount and radius denote the intensity of the sharp details and the sigma parameter of the Gaussian filter, respectively. After testing various combinations, we find that most resulting images are similar. Therefore, we select a fixed amount ($=2$) and radius ($=5$ pixels) to create the ``unsharp masked'' images.

Here, we employ the same characteristics observed in massive galaxies to identify mergers and non-merging galaxies: tidal features, visible large-scale asymmetry within the main body of the galaxy, and shells, as adopted in literature and our previous works (e.g., \citealt{2024MNRAS.529..499L,2025ApJ...986...66T}). We show examples of different types of mergers in Figure \ref{fig:fig1}, with additional merging galaxies available in Figure \ref{fig:fig4} in the Appendix \ref{sec:B1}. Our focus is solely on classifying host galaxies into mergers and non-merging galaxies, because the image depth hampers us from a more refined classification.

\begin{figure}[t]
    \centering
    \subfigure[]{
        \includegraphics[width=0.45\textwidth]{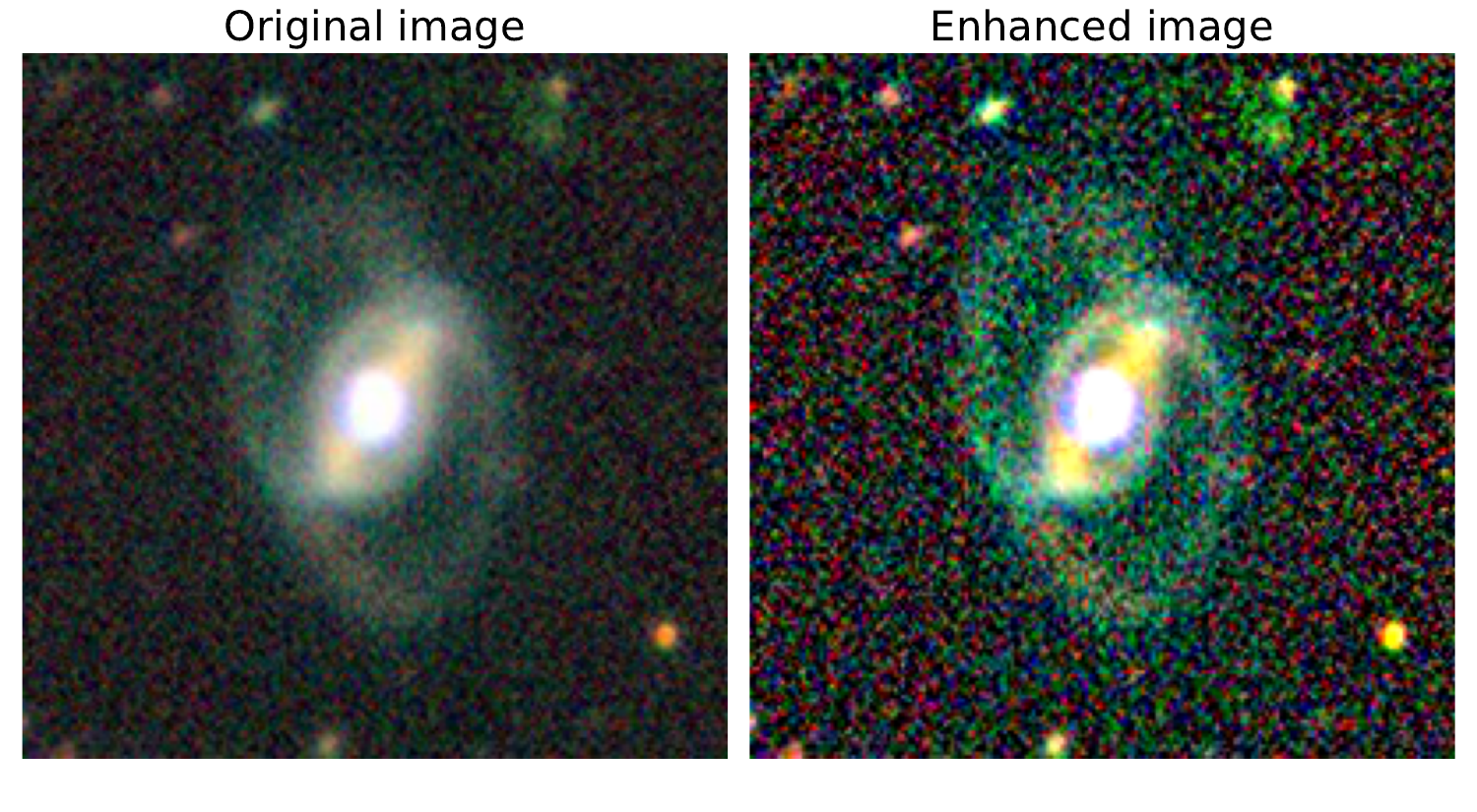}
    }
    \subfigure[]{
        \includegraphics[width=0.45\textwidth]{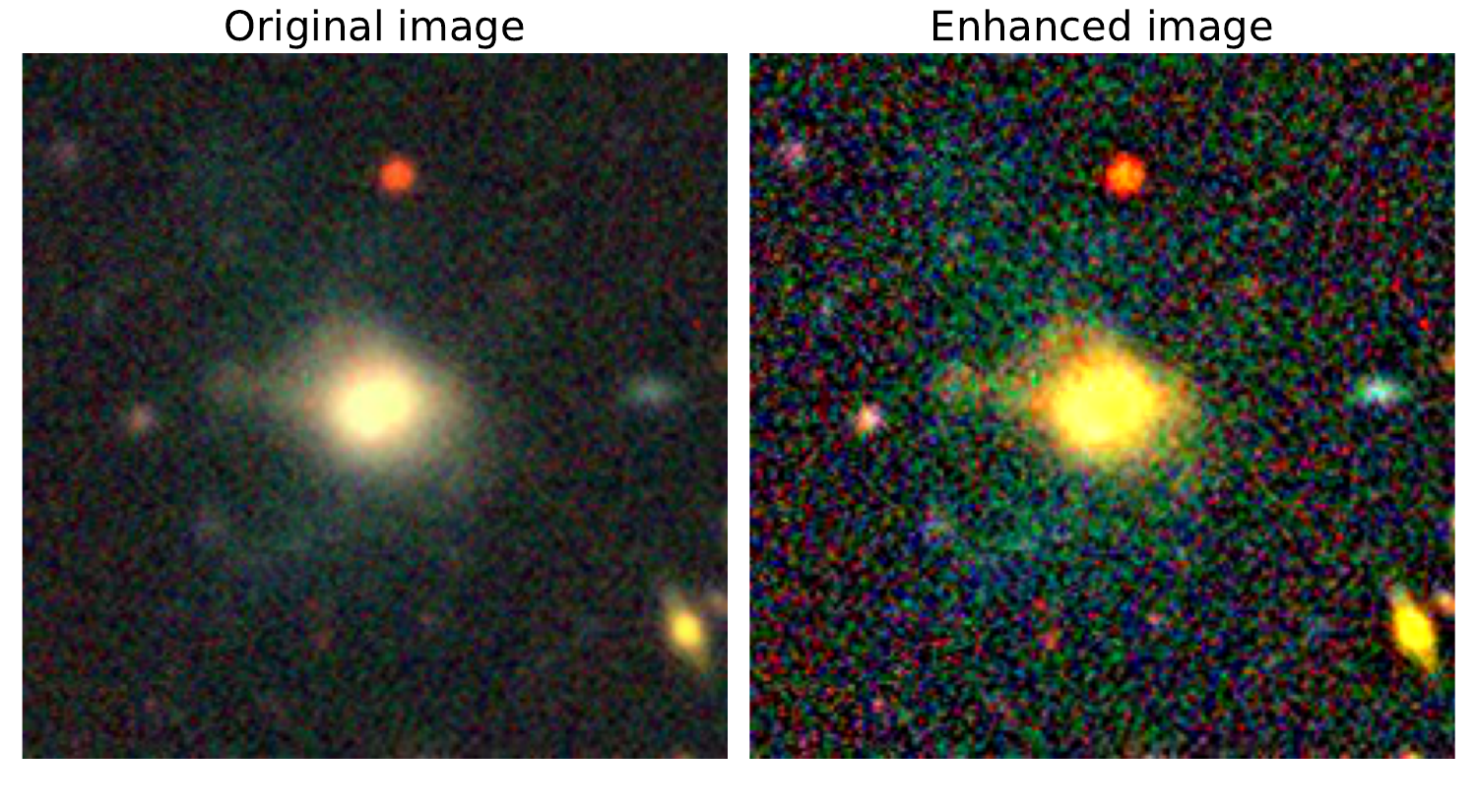}
    }
    \subfigure[]{
        \includegraphics[width=0.45\textwidth]{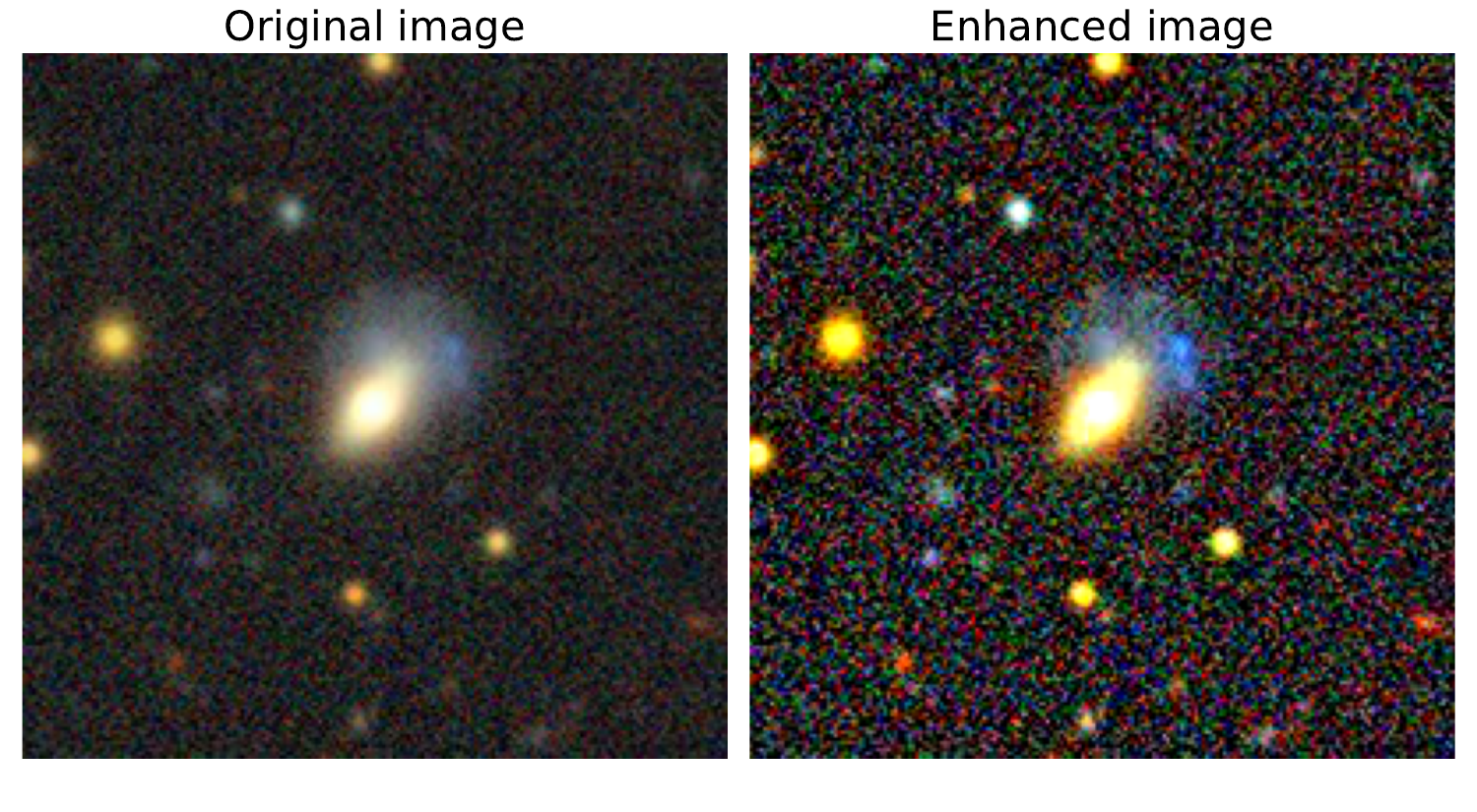}
    }
    \subfigure[]{
        \includegraphics[width=0.45\textwidth]{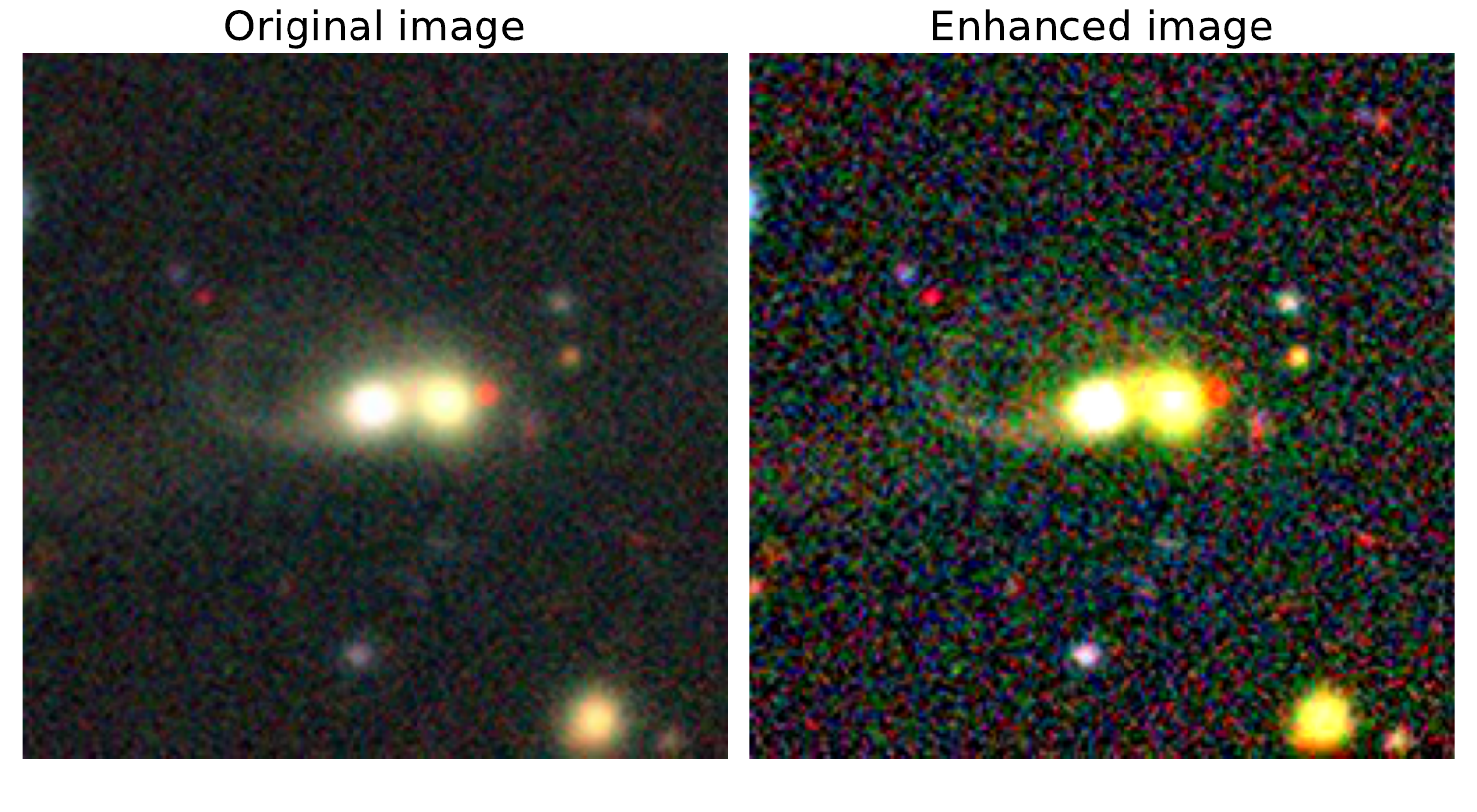}
    }
    \caption{Examples of mergers among CL-AGN host galaxies via visual Classification.
    (a) contains shell structures in its outer region (ID 21); (b) shows a visible large-scale asymmetry within the main body of the galaxy (ID 25); (c) displays clear tidal features (ID 38); (d) shows two cores and an irregular shape (ID 45). }
\label{fig:fig1}
\end{figure}

\section{RESULTS AND DISCUSSION} \label{sec:sec3}
\subsection{Structural Parameters} \label{sec:sec3.1}

In Table \ref{tab:tab1}, we show the mean and median values of the morphological parameters for CL-AGN hosts. For the concentration (\textit{C}), the total sample of CL-AGN host galaxies shows a mean value of $\langle C \rangle=3.3\pm0.4$, close to that of late-type spirals (Sc and Sd; $\langle C \rangle = 3.1\pm0.4$) as reported by \cite{2003ApJS..147....1C}. Additionally, the scatter is also similar. However, CL-AGN hosts have a mean $\langle A \rangle=0.06\pm0.05$, consistent with early-type spirals ($\langle A \rangle = 0.07\pm0.04$; \citealt{2003ApJS..147....1C}), but smaller than late-type spirals ($\langle A \rangle=0.15\pm0.06$) and irregulars ($\langle A \rangle=0.17\pm0.10$) by a factor of more than 2. 

Furthermore, the smoothness parameter ($\langle S \rangle=0.01\pm0.02$) for our CL-AGN hosts is only slightly greater than that of elliptical galaxies ($\langle S \rangle=0.00\pm0.04$), but $8-40\times$ smaller than those of spirals (Sa and Sb: $\langle S \rangle=0.08\pm0.08$; Sc and Sd: $\langle S \rangle=0.29\pm0.12$), and irregulars ($\langle S \rangle=0.40\pm0.20$). As argued in \cite{2003ApJS..147....1C}, \textit{S} can be used to trace a galaxy's star formation properties. It is known that massive ellipticals have little ongoing star formation, while spiral galaxies show modest star-forming activities. Therefore, our results might suggest that the CL-AGN hosts generally contain little ongoing star formation, consistent with the scenario that intense bursts of accretion activity seem to happen more frequently when the supply of cold gas for star formation and accretion decreases \citep{2014MNRAS.444.3078G,2021ApJ...907L..21D}. By contrast, literature works (e.g., \citealt{2021ApJ...907L..21D,2025ApJ...989..101V}) have found that a considerable fraction of their CL-AGN hosts are located in the star-forming main sequence as well as in the so-called ``Green Valley''. This contradiction may be caused by the fact that the smoothness parameter can not be used to identify systems with low to moderate and/or galaxy-spread SF activities. We will return to this point in Section \ref{sec:sec3.3}.
 
In addition to the \textit{CAS} parameters, we also measure \textit{G} and $M_{20}$ to describe the uniformity of the light distribution and the prominent structures within the galaxy, respectively. Our CL-AGN host galaxies show a mean value of $\langle G \rangle=0.54\pm0.04$, similar to early-type spirals ($\langle G \rangle=0.54\pm0.05$) from \cite{2004AJ....128..163L}, and a slightly larger $\langle M_{20} \rangle=-2.0\pm0.2$ (early-type spirals: $\langle M_{20} \rangle=-2.11\pm0.38$). 

Therefore, CL-AGN hosts show complicated structural parameters that differ from any single-type of galaxy. They generally have shallow light profiles comparable to late-type spirals, but are more symmetric (similar to early-type spirals). Furthermore, they show clumpiness similar to ellipticals but much less than spirals. These results seem to suggest that our CL-AGN host galaxies, in general, have a smooth appearance accompanied by weak/modest perturbations. In the following section, we will further explore whether there exist weak/modest signatures of merging/interacting processes (such as stellar shells, faint tidal tails) in these systems with the assistance of our visual classification in the following section.

\subsection{Visual Classification}\label{sec:sec3.2}

We employ the criteria outlined in Section \ref{sec:sec2.4} to classify 63 CL-AGN host galaxies into mergers and non-merging galaxies. As shown in Table \ref{tab:tab1}, we identify that eighteen out of 63 objects show signatures of interacting/merging processes. Therefore, the merging fraction, $f_{\mathrm{merger}}$, is $(28.6\pm5.6)\%$, about 3 times higher than $f_{\mathrm{merger}} = (9.8\pm3.0)\%$ for a sample of 102 AGN-hosts ($0<z<0.3$) selected based on optical variability, estimated by \cite{2022ApJ...925..157Z} using \textit{G}$-M_{20}$ diagnostics and visual classification. Meanwhile, low-redshift galaxies have $f_{\mathrm{merger}}$ in the range of $\sim$1\%$-$6\% \citep{2010MNRAS.401.1043D,2018MNRAS.475.1549M,2021RNAAS...5..144T}. Thus, the much higher $f_{\mathrm{merger}}$ within our sample of CL-AGN host galaxies suggests that mergers/interactions may play an important role in CL phenomena. As suggested by \cite{2023ApJ...956..137W}, such processes can intermittently supply material to the central black holes, resulting in alternating stages of ``feast'' and ``famine''.

We also show the original and enhanced images of all the mergers in our CL-AGN hosts in Figure \ref{fig:fig1} and \ref{fig:fig4}. It is interesting to note that ten out of the 18 merging systems show potential shell structures. Such a high fraction of shell structures is also seen in \cite{2019ApJ...876...75C} (though subject to a large uncertainty of small sample sizes), who identified 3 objects showing shell structures among their 4 CL quasar host galaxies. However, \cite{2018ApJ...866..103K} found that only $(17.8\pm1.1)\%$ of 1201 tidal feature galaxies host shell structures, which is significantly lower than the fraction ($55.6\pm11.1\%$) found in our CL-AGN hosts. The shell structures may be a result of merger events with stellar mass ratios $\gtrsim$1:10, \citep[i.e., intermediate-mass mergers and major mergers;][]{2018MNRAS.480.1715P}, and may have long survival times of $\sim3-4$ Gyr \citep{2019A&A...632A.122M}. 

\subsection{Comparison with NCL-AGN Host Galaxies} \label{sec:sec3.3}

\citetalias{2022ApJ...925...70Z} used \textit{i}-band images from Pan-STARRS to derive the non-parametric coefficients, including \textit{C} and \textit{A}, for low redshift ($0.04<z<0.15$) type 1 (AGN1s) and type 2 (AGN2s) AGN-host galaxies. They considered an extended source to be a possible companion galaxy if it lies within $2r_{\mathrm {petro}}$ \citep{1976ApJ...209L...1P} of the main target and contains at least 25\% of the main target's flux. Such systems are considered as major merger candidates. Here we compare our results with those from \citetalias{2022ApJ...925...70Z} to explore whether there is any difference in morphology between CL- and NCL-AGN host galaxies.

Before the comparison, We removed the duplicates (2 AGN1s and 15 AGN2s that appear in both our sample and theirs) from their sample, resulting in a sample of NCL-AGNs comprising 243 AGN1s and 4499 AGN2s. While it might still miss some potential CL-AGNs in \citetalias{2022ApJ...925...70Z} sample, we think that the missed CL-AGNs only occupy a small fraction and thus will not significantly affect the average properties of the NCL-AGN sample. This is because the number of CL-AGNs remains quite limited in comparison to AGNs \citep{2024ApJ...966..128W}, roughly 0.3\% for $z<0.35$ AGNs over $5 \sim 20$ years. For NCL-AGN hosts \citetalias{2022ApJ...925...70Z}, $f_{\mathrm{merger}}$ is $(3.1\pm0.3)\%$ (3.7\% for AGN1s and 3.0\% for AGN2s), which are lower than our CL-AGN hosts by a factor of more than 8. However, this difference might be caused by the different methods used to select merger candidates, because relying solely on major mergers could lead to an underestimation of the true merger fraction. Thus, we construct three control samples with different selection criteria from the NCL-AGN catalog for our CL-AGN sample, and perform the same visual classification as described in Section \ref{sec:sec2.4}. 

To select the control sample sources, we first define the difference of redshifts as, 
\begin{equation}
    \Delta z =  \left |  \frac{z_\mathrm{CL}-z_\mathrm{NCL}}{z_\mathrm{CL}}   \right |,
\end{equation}
where $z_\mathrm{CL}$ and $z_\mathrm{NCL}$ represent the redshift of CL-AGNs and NCL-AGNs, respectively. In addition, we will construct three control samples by matching black hole mass, stellar mass of the host galaxies, and AGN strength (i.e., the [OIII] luminosity, $L_\mathrm{[OIII]}$), respectively, to our CL-AGNs. For the control samples, these parameters are adopted from \citetalias{2022ApJ...925...70Z}.

For control \textbf{Sample I}, due to the limited size of the AGN1s, we matched each CL-AGN to only one object from AGN1s based on (1) $\Delta z \leq 0.2 $ and (2) black hole mass ($M_{\bullet}$) from \cite{2025ApJS..278...28G}:
\begin{equation}
    \Delta M_{\bullet} =  \left | \log M_\mathrm{\bullet,CL}- \log M_\mathrm{\bullet,NCL}  \right |  \leq 0.3,
\end{equation}
where $M_\mathrm{\bullet,CL}$ ($M_\mathrm{\bullet,NCL}$) represents the black hole mass of CL-AGNs (NCL-AGNs). Only 58 CL-AGNs are found to have a counterpart, and these 58 CL-AGNs have a slightly higher merger fraction ($f_{\mathrm{merger,CL}} = 31.0\pm6.0\%$). 

Among control \textbf{Sample I} (58 objects), seven show merging signatures. Therefore, the merger fraction of black hole mass-matched NCL-AGNs is $f_{\mathrm{merger,NCL}} = (12.1\pm4.3)\%$, $>$2 times lower than that of our CL-AGNs. It is interesting to note that five out of 7 NCL-AGN mergers ($71.4\pm15.5\%$) show shell structures, which is a bit higher than our CL-AGNs ($55.6\pm11.1\%$).

We also constructed control \textbf{Sample II} by selecting two matched objects (to increase the sample size and thus reduce uncertainties) from the AGN2s for each CL-AGN, using constraints of (1) $\Delta z \leq 0.1 $ and (2) stellar mass ($M_{\star}$) from the GALEX-SDSS-WISE Legacy Catalog 2 (GSWLC-2; \citealt{2018ApJ...859...11S}) for both samples:
\begin{equation}
    \Delta M_{\star} =  \left |  \log M_\mathrm{\star,CL}-\log M_\mathrm{\star,NCL}  \right |  \leq 0.1,
\end{equation}
where $M_\mathrm{\star,CL}$ and $M_\mathrm{\star,NCL}$ represent the stellar mass of CL-AGN and NCL-AGN host galaxies, respectively. We found that forty-nine CL-AGNs have $M_{\star}$ measurements in the GSWLC-2 catalog, and these CL-AGNs have $f_{\mathrm{merger,CL}} =(24.5\pm6.0)\%$. 

Control \textbf{Sample II} consists of 98 sources, out of which thirteen are visually classified as mergers, corresponding to $f_{\mathrm{merger,NCL}} = (13.3\pm3.5)\%$. This merger fraction is about 2 times smaller than our CL-AGN sample. There are 6 out of 13 NCL-AGN mergers ($46.2\pm12.9\%$) have shell structures and the fraction is close to our CL-AGNs ($58.3\pm13.2\%$).

To select counterparts with similar AGN strength to our CL-AGNs, we matched two NCL-AGNs to each CL-AGNs by constraining (1) $\Delta z \leq 0.1 $ and (2) the [OIII]\,$\lambda$5007 line luminosity, as it is a good tracer of AGN activity (see; e.g., \citealt{2003MNRAS.346.1055K}), with $\Delta \log L_\mathrm{[OIII]} =  \left |   \log L_\mathrm{[OIII],CL}- \log L_\mathrm{[OIII],NCL}  \right |  \leq 0.2$. Here $L_\mathrm{[OIII],CL}$ and $L_\mathrm{[OIII],NCL}$ represent the [OIII]\,$\lambda$5007 line luminosity of CL-AGNs and NCL-AGNs, respectively, and are computed with the [OIII]\,$\lambda$5007 fluxes from \cite{2025ApJ...994..203S} and \citetalias{2022ApJ...925...70Z}. 

For this $L_\mathrm{[OIII]}$-matched control sample (\textbf{Sample III}), we identified that seventeen out of 126 objects are mergers, i.e., $f_{\mathrm{merger,NCL}} = (13.5\pm3.1)\%$, a fraction similar to control \textbf{Sample I} and \textbf{Sample II}, but about half of the CL-AGNs ($28.6\pm5.6\%$). Among these 17 mergers, nine objects exhibit a shell structure ($52.9\pm11.5\%$), similar to our CL-AGNs ($55.6\pm11.1\%$). 

Comparison across all three control samples consistently demonstrates that the merger fraction of our CL-AGN sample is $\sim$2$\times$ as high as that of the NCL-AGNs. This result indicates that mergers/interactions may play an important role in the CL phenomenon. In addition, the average fraction of sources showing shell structures for these three control samples is 56.8\%, consistent with CL-AGNs, which suggests that intermediate-mass mergers and major mergers are prevalent in the merging events of AGN-hosts based on simulations (\citealt{2018MNRAS.480.1715P, 2019A&A...632A.122M}). However, future studies using a much larger sample and/or deeper images are needed to draw a solid conclusion.

\begin{figure}[t]   
    \centering
    \includegraphics[width=0.8\textwidth]{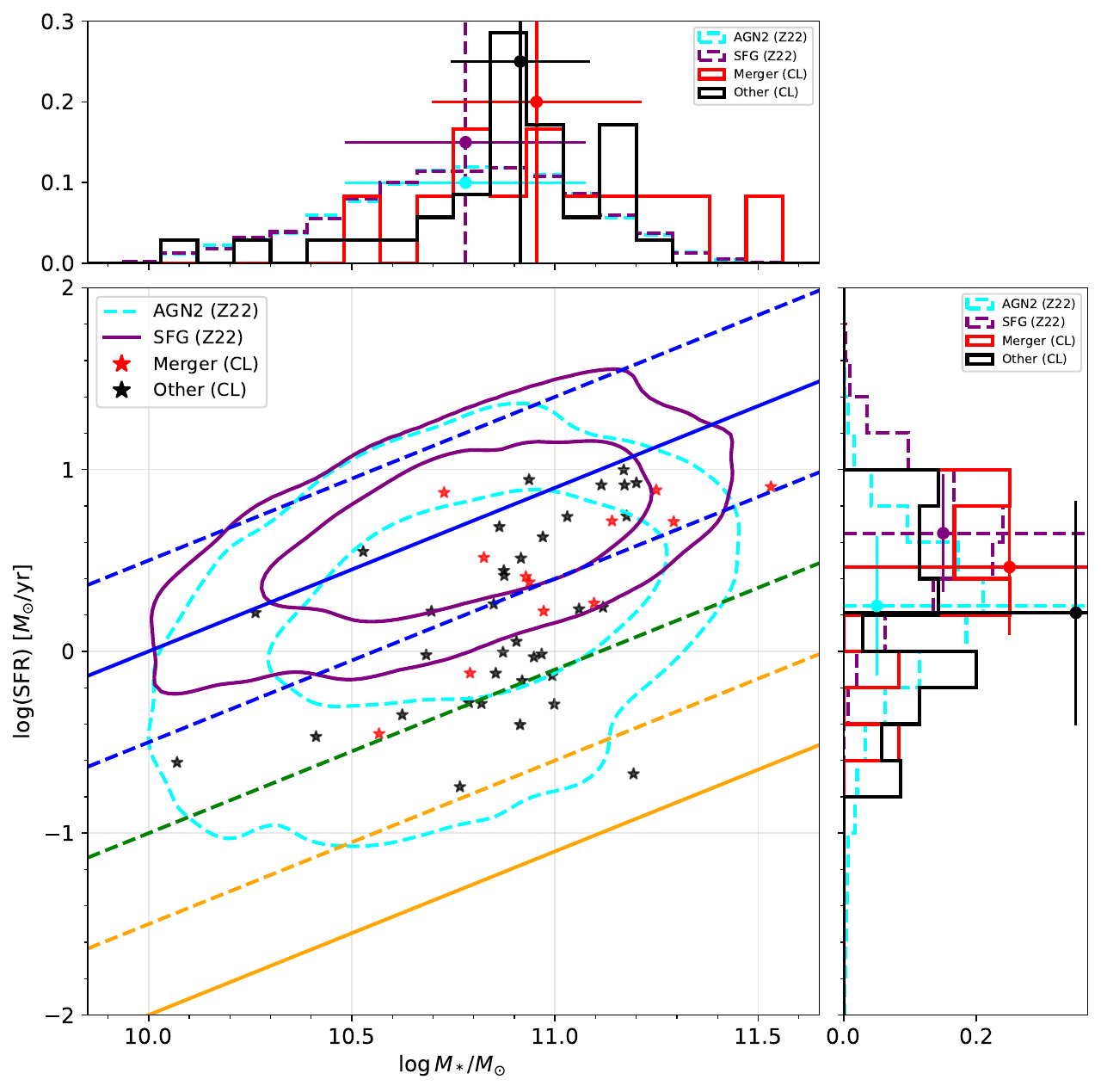}
    \caption{SFR vs. $M_{\star}$ for CL-AGNs and NCL-AGNs from \citetalias{2022ApJ...925...70Z}. The solid blue line is the star-forming main sequence \citep{2010ApJ...721..193P}. Dashed lines are spaced $1\sigma$ from the main sequence \citep{2017ApJ...850...22L}. The Green Valley lies $1-3\sigma$ below the main sequence, between the lower blue dashed line and the orange dashed line. Below the orange dashed line lie quiescent galaxies. Red and black stars denote mergers and other galaxies in the CL-AGN sample, respectively; green squares and gray dots denote mergers and other galaxies in the NCL-AGN sample, respectively. The cyan and purple lines illustrate the 68\% and 95\% contours for AGN2s and SFGs from \citetalias{2022ApJ...925...70Z}, respectively. Distributions (shown by the histograms) of the parameters and median values (shown by the lines, together with their associated uncertainties) are also plotted on the sides of the main panel.}
    \label{fig:fig6}
\end{figure}

To further investigate whether these mergers are linked to the star formation properties of CL-AGN hosts, we examine their SFR-$M_{\star}$ relation in Figure \ref{fig:fig6}, which plots the SFR versus $M_{\star}$ for our 47 turn-off CL-AGNs, and the NCL-AGN2 and star-forming (SFGs) samples from \citetalias{2022ApJ...925...70Z}. Here the SFR and $M_{\star}$ are all adopted from the GSWLC-2 catalog, which provides derived galactic parameters obtained by fitting the spectral energy distribution (SED) from the ultraviolet to the mid-infrared bands \citep{2018ApJ...859...11S}. We do not use the H$\beta$ (or H$\alpha$) line to measure SFRs to avoid the contamination from AGNs, since most of our sources are classified as Seyfert galaxies \citep{2025ApJS..278...28G}.

As shown in Figure \ref{fig:fig6}, the median SFR of mergers in our CL-AGN sample is about 2 times higher than that of non-mergers, indicating that star formation activity is enhanced in merging systems. It is also demonstrated in Figure \ref{fig:fig6} that CL-AGNs generally follow a similar distribution to AGN2s in the SFR-$M_{\star}$ plot, consistent with \citep{2025ApJ...989..101V}, who reported that turn-off CL-AGNs show similar SF properties to Type 1 and Type 2 Seyfert hosts at $z<0.15$.

Furthermore, we find that $\sim$49\% of our CL-AGN host galaxies fall within the $\pm1\sigma$ region of the SF main sequence, which is similar to the $\sim 40-50\%$ SF fraction in \cite{2025ApJ...989..101V}. However, unlike SFGs which are distributed on both sides of the main sequence (purple contours of Figure \ref{fig:fig6}), our CL-AGNs are located mainly below the (blue) line of the SF main sequence. We also note that about half ($48.9\pm7.1\%$) of CL-AGNs are located in the Green Valley, which agrees within $3\sigma$ uncertainties with those given in previous works (e.g.,  \citealt{2021ApJ...907L..21D}: $75.0\pm14.3\%$; \citealt{2025ApJ...989..101V}: $29.2\pm9.0\%$) within the same redshift ($z<0.15$).

Therefore, the overwhelming majority of our turn-off CL-AGN hosts exhibit ongoing SF activities at different levels, which differs from the result based on the smoothness parameter in Section \ref{sec:sec3.1}. This discrepancy may arise from the fact that (1) the measured $S$ tends to be lower at higher redshifts (\citealt{2003ApJS..147....1C}); and/or (2) it is hard to employ $S$ to identify the SF properties when SF activities spread widely in a galaxy, especially when the SFR is relatively small.

\begin{table}[t]
\movetableright=-25mm
\caption{Comparison of Merger and Shell Fractions Between CL-AGNs and Their NCL-AGN Counterparts.}
\label{tab:tab2}
\centering{
\begin{tabular}{ccccc}\\\hline \hline
Sample     & Type    & Number & Merger Fraction  & Shell Fraction  \\ \hline
Sample I   & CL-AGN  & 58     & $(31.0\pm6.0)\%$ & $(55.6\pm11.1)\%$ \\
           & NCL-AGN & 58     & $(12.1\pm4.3)\%$ & $(71.4\pm15.5)\%$ \\\hline
Sample II  & CL-AGN  & 49     & $(24.5\pm6.0)\%$ & $(58.5\pm13.2)\%$ \\
           & NCL-AGN & 98     & $(13.3\pm3.5)\%$ & $(46.2\pm12.9)\%$ \\\hline
Sample III & CL-AGN  & 63     & $(28.6\pm5.6)\%$ & $(55.6\pm11.1)\%$ \\
           & NCL-AGN & 126    & $(13.5\pm3.1)\%$ & $(52.9\pm11.5)\%$ \\\hline
\end{tabular}
}
\tablecomments{The $1\sigma$ uncertainty for each fraction is the same as Table \ref{tab:tab1}.}
\end{table}

\subsection{Comparison of turn-on and off objects} \label{sec:sec3.4}

\begin{figure}[ht!]
\centering
\subfigure[Concentration vs Asymmetry]{
       \includegraphics[width=0.45\textwidth]{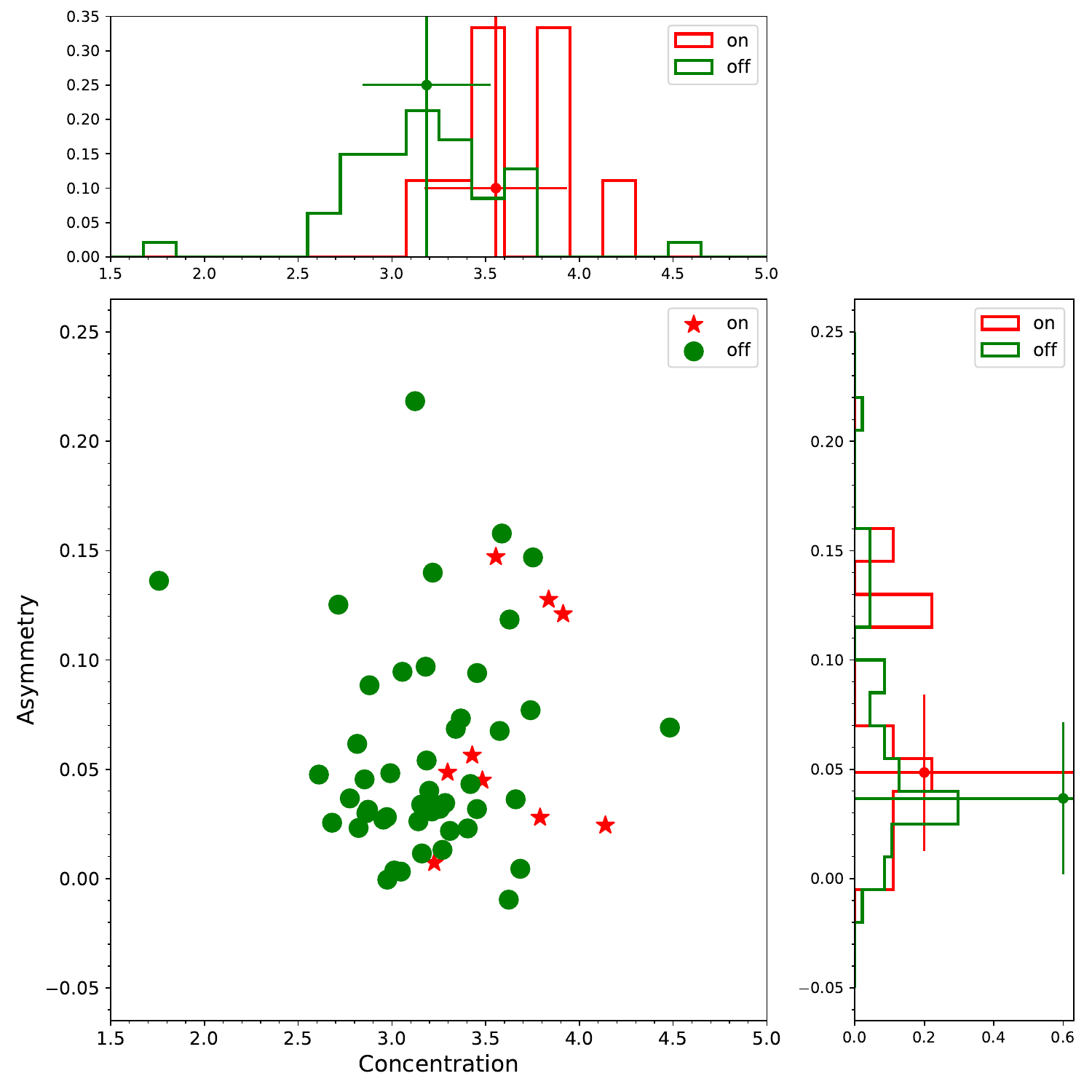}
       }
\subfigure[Concentration vs Clumpiness]{
       \includegraphics[width=0.45\textwidth]{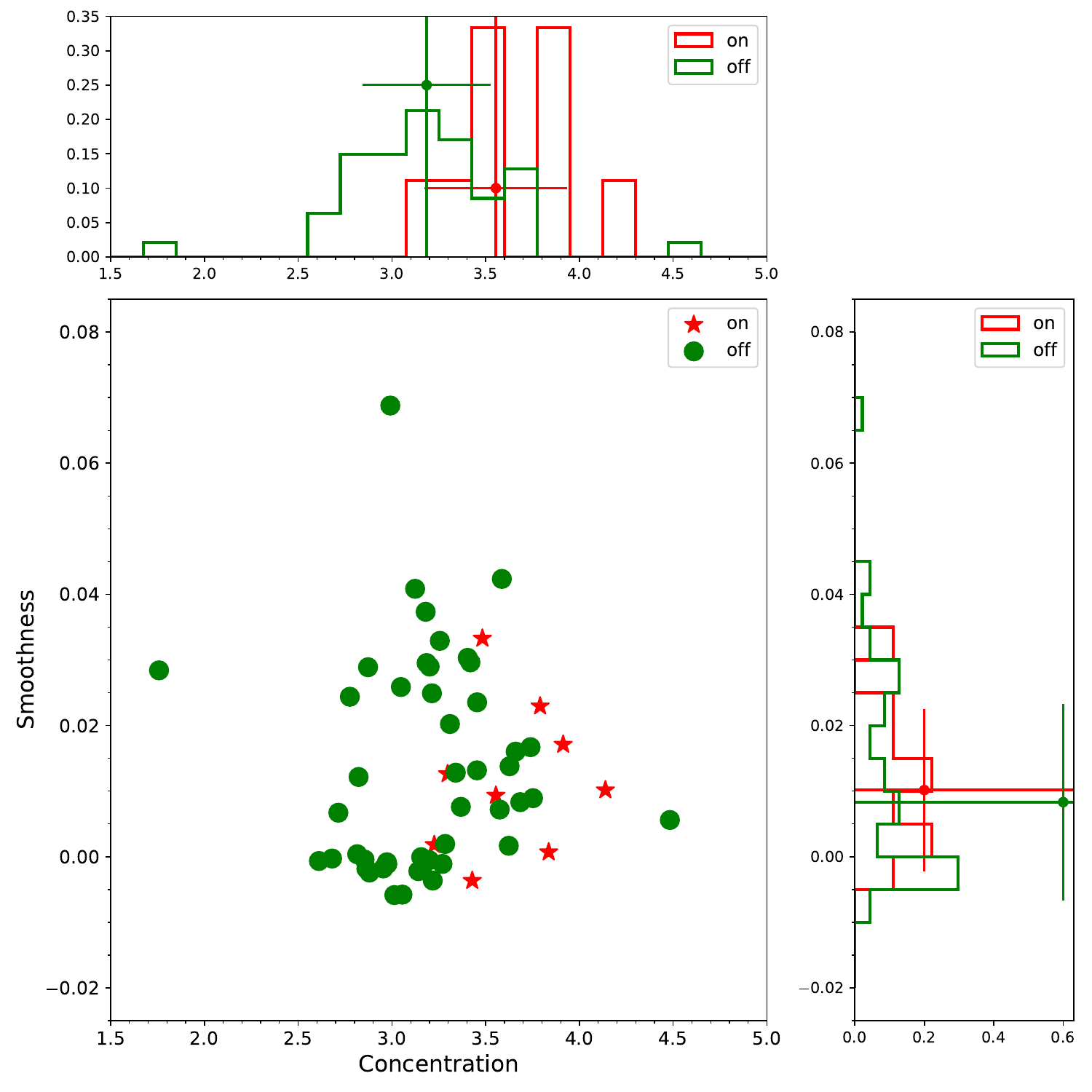}
       }
\subfigure[Asymmetry vs Clumpiness]{
       \includegraphics[width=0.45\textwidth]{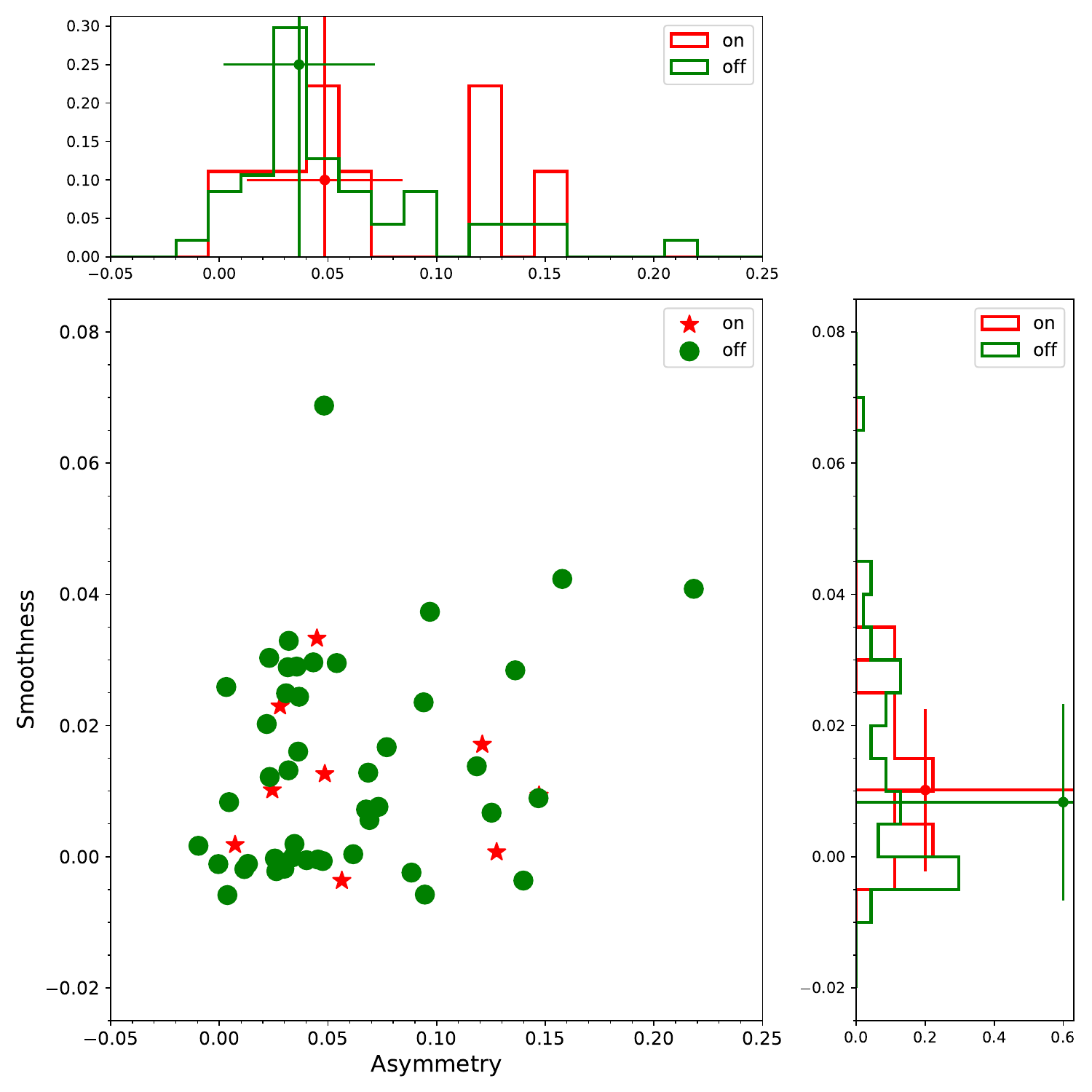}
       }
\subfigure[Gini vs $M_{20}$]{
       \includegraphics[width=0.45\textwidth]{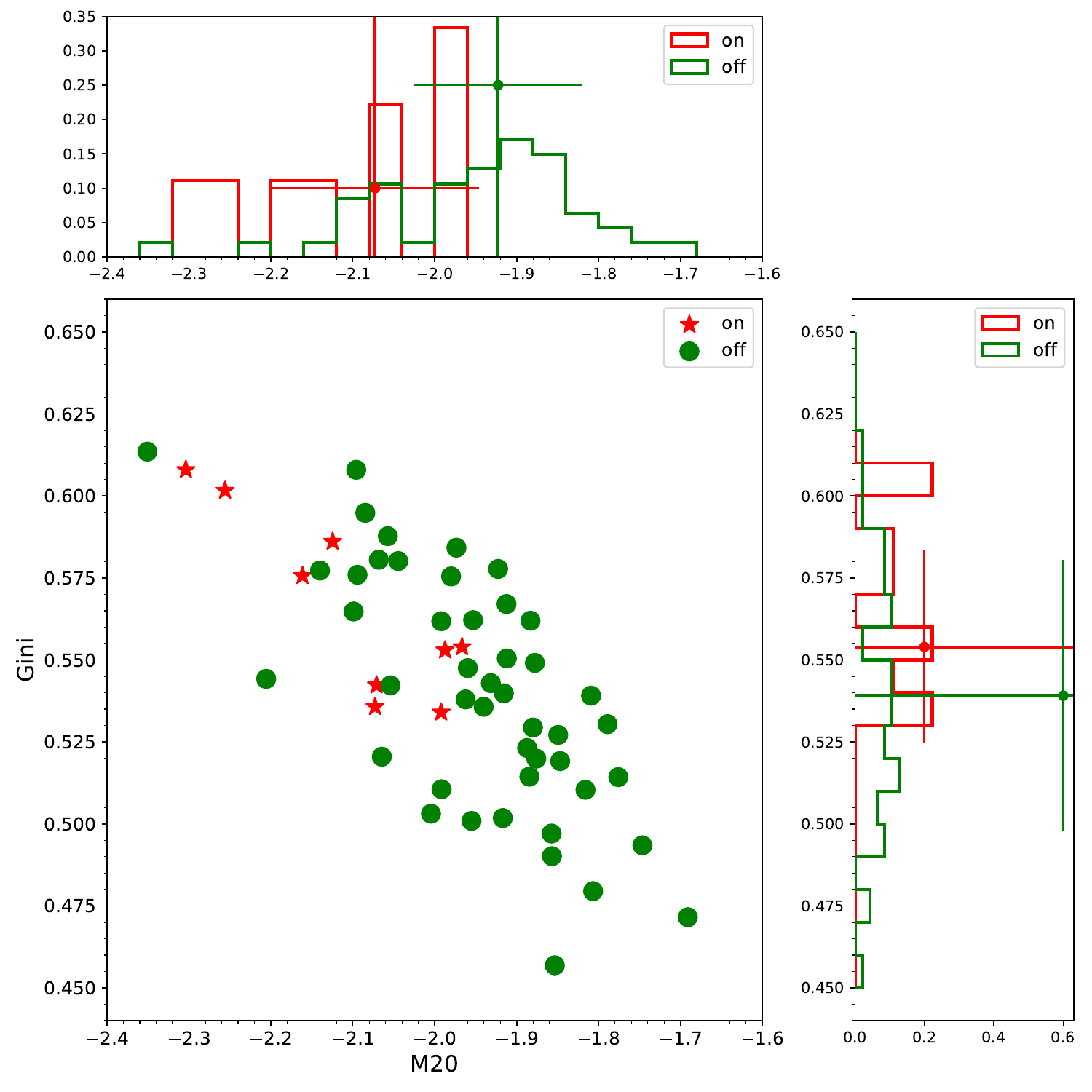}
       }
\caption{Concentration vs Asymmetry (a),  Concentration vs Clumpiness (b), Asymmetry vs Clumpiness (c), and Gini vs $M_{20}$ (d) for ``turn-on'' (on, red stars) and ``turn-off'' (off, green circles) CL-AGN hosts. Distributions (shown by the histograms) of parameters and median values (shown by the lines, together with their associated uncertainties) are also plotted on the sides of the main panel. }
\label{fig:fig3}
\end{figure}

For the purpose of our work, the bright nuclei of turn-on CL-AGNs may affect the measurements to some extent. However, we do not know which CL-AGN is in the turn-on state since the imaging and spectroscopic data were not taken simultaneously. In the following paragraph, we will make a rough estimate of the stage for these CL-AGNs using the observational time of each imaging data.   

The intervals between the two spectral epochs for our CL-AGNs range from approximately 3000 to 8000 days. The Modified Julian Date (MJD) for each imaging epoch from DESI DR10 falls within these two spectral epochs, with the small interval exceeding approximately 1000 days. Consequently, the stage (imaging data) of a CL-AGN is assigned to the spectroscopic (DESI/SDSS) one that has the minimum interval between these two observations. Here, we introduce a parameter $x$ to quantify the distance between the imaging MJD and the two spectral MJDs:
\begin{equation}
     x = \frac{\mathrm{MJD}_\mathrm{DIm}-\mathrm{MJD}_\mathrm{SSp}}{ \mathrm{MJD}_\mathrm{DSp}-\mathrm{MJD}_\mathrm{SSp}},
\end{equation}
where $\mathrm{MJD}_\mathrm{DIm}$ represent the MJD of the image from DESI DR10. $\mathrm{MJD}_\mathrm{DSp}$ and $\mathrm{MJD}_\mathrm{SSp}$ are the MJDs of the spectral epochs of DESI and SDSS, respectively. If $x \ge 0.5$, the image is thought to have the same state as shown in the DESI spectrum; otherwise, it is assigned the state determined by the SDSS spectrum. 

Based on the $x$ values and CL-AGN stages provided in \cite{2025ApJS..278...28G} catalog, we find that the ratio of turn-on to turn-off CL-AGNs is $\sim 1:5$. The mean and median of the corresponding morphological parameters, along with their respective counts and fractions, are also presented in Table \ref{tab:tab1}. To obtain a more comprehensive analysis, we show the scatter plots of \textit{CAS} and $G-M_{20}$ in Figure \ref{fig:fig3}, alongside histograms and medians displayed on either side. 

As shown in Table \ref{tab:tab1} and Figure \ref{fig:fig3}, indeed, the host galaxies of turn-on CL-AGNs show some differences in \textit{C}, \textit{G}, and $M_{20}$, all of which are correlated with concentration. The mean values of \textit{C} and \textit{G} are larger in turn-on ($\langle C \rangle=3.6\pm0.3$ and $\langle G \rangle=0.57\pm0.03$) than turn-off CL-AGN hosts ($\langle C \rangle=3.2\pm0.4$ and $\langle G \rangle=0.54\pm0.04$). This suggests that AGN in the turn-on stage may have a significant influence on a galactic scale, characterized by a more luminous central region. This luminous central region reduces the relative contribution of other bright structures to the galaxy's overall light distribution, which may explain the noticeably smaller $M_{20}$ values observed in these turn-on ($\langle M_{20} \rangle=-2.1\pm0.1$) than turn-off ($\langle M_{20} \rangle=-1.9\pm0.2$) CL-AGN hosts. 

We also calculate $f_{\mathrm{merger}}$ for these two stages, and find $f_{\mathrm{merger}} = (25.5\pm6.3)\%$ and $f_{\mathrm{merger}} = (30.0\pm13.5)\%$ for the turn-off and -on stages, respectively. The $>$2$\times$ uncertainty for the turn-on stage can be attributed to its much smaller sample size. The slight difference of $f_{\mathrm{merger}}$ for these two stages of CL-AGN hosts suggests that the timescale for mergers is significantly longer than that for the CL phenomenon. As discussed in Section \ref{sec:sec3.3}, CL-AGN host galaxies are preferentially located in the star-forming main sequence and the Green Valley, and star formation activity is enhanced in mergers. Since most mergers in our sample are post-merger systems, this suggests they are in the late ``feast'' stage rather than the ``famine'' stage. Specifically, cold gas - though partially consumed - is still abundant enough to fuel enhanced star formation, even as the system has already begun moving toward the Green Valley. Nevertheless, it is important to note that our classification of CL-AGN stages based on MJDs is relatively naive, which may lead to some mis-classifications. Therefore, future works using simultaneous observations of spectroscopic and imaging data are needed to obtain robust results.

\section{SUMMARY}\label{sec:sec4}
In this study, we select a subsample of 63 spectroscopically identified CL-AGNs with $z<0.15$ from  \cite{2025ApJS..278...28G} catalog, using DESI DR10 image data to perform non-parametric measurements and conduct a visual classification. We also compare the results between CL-AGN and NCL-AGN hosts, as well as between turn-on and turn-off stages. Our main results are:  

\begin{enumerate}
    \item The morphology of CL-AGN host galaxies is complicated and differs from any single-type of galaxy, with concentration similar to late-type spirals, asymmetry close to early-type spirals, and smoothness consistent with ellipticals.
    \item We identify that CL-AGN hosts consist of approximately 29\% merging/interacting systems, significantly higher than NCL-AGN hosts, indicating that mergers/interactions may play an important role in driving the CL phenomenon.
    \item Stellar shells are prevalent in merging systems of AGN-hosts (56\%), suggesting a significant role of intermediate-mass mergers and major mergers.
\end{enumerate}

\begin{acknowledgments}
The authors thank the anonymous referee for valuable comments/suggestions. This work is supported by the China Manned Space Program with grant No. CMS-CSST-2025-A08 and the National Key R\&D Program of China (No. 2021YFA1600404). W.J.G. acknowledges financial support from National Natural Science Foundation of China (NSFC; grant No. 12503019). All the authors acknowledge the work of the Dark Energy Spectroscopic Instrument (DESI) team. The DESI Legacy Imaging Surveys consist of three individual and complementary projects: the Dark Energy Camera Legacy Survey (DECaLS), the Beijing-Arizona Sky Survey (BASS), and the Mayall \textit{z}-band Legacy Survey (MzLS). DECaLS, BASS, and MzLS together include data obtained, respectively, at the Blanco telescope, Cerro Tololo Inter-American Observatory, NSF’s NOIRLab; the Bok telescope, Steward Observatory, University of Arizona; and the Mayall telescope, Kitt Peak National Observatory, NOIRLab. NOIRLab is operated by the Association of Universities for Research in Astronomy (AURA) under a cooperative agreement with the National Science Foundation. Pipeline processing and analyses of the data were supported by NOIRLab and the Lawrence Berkeley National Laboratory (LBNL). Legacy Surveys also uses data products from the Near-Earth Object Wide-field Infrared Survey Explorer (NEOWISE), a project of the Jet Propulsion Laboratory/California Institute of Technology, funded by the National Aeronautics and Space Administration. Legacy Surveys was supported by: the Director, Office of Science, Office of High Energy Physics of the U.S. Department of Energy; the National Energy Research Scientific Computing Center, a DOE Office of Science User Facility; the U.S. National Science Foundation, Division of Astronomical Sciences; the National Astronomical Observatories of China, the Chinese Academy of Sciences and the Chinese National Natural Science Foundation. LBNL is managed by the Regents of the University of California under contract to the U.S. Department of Energy. 
Software: \texttt{statmorph} \citep{2019MNRAS.483.4140R}; \texttt{Photutils} \citep{2020zndo...4049061B}; \texttt{SEXtractor} \citep{1996MNRAS.279L..47A}; \texttt{Unsharp-masking} \citep{2014PeerJ...2..453V}.
\end{acknowledgments}

\appendix
\section{Non-parameter measurements of Cl-AGNs}\label{sec:A1}

\begin{table}[ht!]
\movetableright=-25mm
\centering
\tiny
\caption{Non-parameter measurements of CL-AGNs}
\label{tab:tab3}
\begin{tabular}{ccccccccccc}\\\hline \hline
ID &RA        & DEC      & Z       & merger & Concentration & Asymmetry    & Smoothness   & Gini        & M20          & type \\\hline
1  & 3.44031   & 29.35822 & 0.09466 & m      & 3.283991571            & 0.03456237             & 0.001944732            & 0.550465568            & -1.912071596           & off  \\
2  & 25.47489  & 1.08485  & 0.10141 & m      & 3.179944039            & 0.096915694            & 0.037339012            & 0.510579107            & -1.991731771           & off  \\
3  & 29.5198   & -0.87275 & 0.08065 & n      & \nodata & \nodata & \nodata & \nodata & \nodata & off  \\
4  & 32.03652  & 0.51637  & 0.08412 & n      & 2.853866034            & 0.045391178            & -0.000424269           & 0.490169993            & -1.85702338            & off  \\
5  & 111.73366 & 41.02668 & 0.12978 & m      & 3.913298885            & 0.121018074            & 0.017103458            & 0.601632094            & -2.2559402             & on   \\
6  & 116.5238  & 22.27831 & 0.08271 & n      & 3.056081661            & 0.094582265            & -0.005773357           & 0.519909001            & -1.876103512           & off  \\
7  & 118.48353 & 26.88994 & 0.13415 & m      & 3.454410868            & 0.094015755            & 0.023541889            & 0.580548268            & -2.068315807           & off  \\
8  & 119.2201  & 29.3284  & 0.06785 & m      & \nodata & \nodata & \nodata & \nodata & \nodata & off  \\
9  & 120.76977 & 22.12613 & 0.12475 & n      & 2.954164183            & 0.026980406            & -0.001784341           & 0.530431429            & -1.789081886           & off  \\
10 & 121.22592 & 21.60753 & 0.10986 & n      & 3.159838396            & 0.01149628             & -0.001877172           & 0.562143787            & -1.953188059           & off  \\
11 & 123.16984 & 7.25794  & 0.0852  & m      & \nodata & \nodata & \nodata & \nodata & \nodata & off  \\
12 & 128.2938  & 4.17685  & 0.10122 & n      & 3.418621831            & 0.043254223            & 0.029632651            & 0.561799847            & -1.992076592           & off  \\
13 & 129.31863 & 3.93503  & 0.06369 & m      & 3.25551797             & 0.031936355            & 0.032911422            & 0.503117603            & -2.004608866           & off  \\
14 & 130.45578 & -0.17542 & 0.12855 & m      & 3.340516768            & 0.068501707            & 0.012816062            & 0.542237665            & -2.053994017           & off  \\
15 & 131.94578 & 39.49406 & 0.12309 & n      & 2.714183328            & 0.125289324            & 0.00671062             & 0.45688417             & -1.853495167           & off  \\
16 & 134.3887  & 14.04818 & 0.13422 & n      & \nodata & \nodata & \nodata & \nodata & \nodata & on   \\
17 & 142.89271 & 0.50784  & 0.07336 & n      & 2.880266289            & 0.088419768            & -0.002414968           & 0.501783874            & -1.916980886           & off  \\
18 & 171.83123 & 4.19923  & 0.0748  & m      & 3.627476044            & 0.118490823            & 0.013791482            & 0.577263395            & -2.140074007           & off  \\
19 & 172.41111 & 21.12883 & 0.10668 & n      & 2.973027783            & 0.028133659            & -0.000844671           & 0.500912122            & -1.955166463           & off  \\
20 & 173.12141 & 3.95809  & 0.09118 & n      & \nodata & \nodata & \nodata & \nodata & \nodata & off  \\
21 & 179.90787 & 65.76015 & 0.12184 & m      & 4.482653681            & 0.069083826            & 0.005600849            & 0.613510462            & -2.350832737           & off  \\
22 & 179.97246 & 5.225    & 0.05924 & n      & \nodata & \nodata & \nodata & \nodata & \nodata & off  \\
23 & 181.47034 & 33.57246 & 0.11562 & n      & 2.610521387            & 0.047579447            & -0.000650576           & 0.471560764            & -1.691146334           & off  \\
24 & 184.07136 & 56.85144 & 0.11107 & n      & 3.213401617            & 0.030710777            & 0.024918435            & 0.537977447            & -1.962343957           & off  \\
25 & 186.30698 & 60.89438 & 0.12429 & m      & 3.296919576            & 0.048523224            & 0.012619334            & 0.552971742            & -1.987440358           & on   \\
26 & 192.52235 & 4.26656  & 0.12345 & n      & 2.99133127             & 0.048201021            & 0.068781192            & 0.523159039            & -1.887296106           & off  \\
27 & 193.51578 & 49.24801 & 0.06713 & n      & 3.660036882            & 0.036266259            & 0.016025602            & 0.575965191            & -2.09430333            & off  \\
28 & 196.9262  & 52.04446 & 0.10301 & n      & 2.822060031            & 0.023169023            & 0.012159344            & 0.527136902            & -1.849331502           & off  \\
29 & 202.3994  & -1.50946 & 0.08217 & n      & 3.40425015             & 0.022940767            & 0.030320867            & 0.575471115            & -1.980058491           & off  \\
30 & 203.8352  & -0.97743 & 0.14287 & m      & \nodata & \nodata & \nodata & \nodata & \nodata & off  \\
31 & 205.87506 & 51.03448 & 0.13182 & n      & 3.185024831            & 0.054015526            & 0.029527252            & 0.567058976            & -1.912405851           & off  \\
32 & 206.08166 & 51.44014 & 0.06292 & n      & 3.752496686            & 0.146894503            & 0.00892876             & 0.544233428            & -2.205734866           & off  \\
33 & 207.49467 & -0.18914 & 0.10202 & m      & 2.775524032            & 0.03667986             & 0.0243884              & 0.479511223            & -1.806747372           & off  \\
34 & 211.97276 & 28.23586 & 0.11227 & n      & 3.309436494            & 0.021805788            & 0.020234549            & 0.547524604            & -1.959635454           & off  \\
35 & 212.49051 & -1.48071 & 0.13558 & n      & 3.482409737            & 0.044920546            & 0.033283017            & 0.535696189            & -2.072906542           & on   \\
36 & 217.56688 & 23.06234 & 0.08103 & n      & 3.739407316            & 0.07702255             & 0.016701981            & 0.59486302             & -2.084775855           & off  \\
37 & 223.77515 & 11.85585 & 0.12465 & n      & 2.975310526            & -0.000485764           & -0.001121546           & 0.514400444            & -1.884529366           & off  \\
38 & 225.23318 & -1.87345 & 0.11109 & m      & 3.217833537            & 0.139920538            & -0.003632646           & 0.539111811            & -1.809224499           & off  \\
39 & 228.3813  & 4.0738   & 0.11111 & n      & 3.140899159            & 0.026218095            & -0.00219845            & 0.535715844            & -1.940312947           & off  \\
40 & 228.4407  & 31.1903  & 0.07205 & n      & 3.574718536            & 0.067527413            & 0.007197332            & 0.587750805            & -2.057296127           & off  \\
41 & 230.6883  & 42.74752 & 0.10962 & n      & 3.19895592             & 0.040160955            & -0.000525371           & 0.539825472            & -1.915742237           & off  \\
42 & 232.87404 & 32.9832  & 0.12597 & n      & 3.428271235            & 0.056383279            & -0.003649506           & 0.553953566            & -1.966714895           & on   \\
43 & 233.29177 & 27.48895 & 0.07207 & n      & 3.155933569            & 0.033823982            & -6.71E-05              & 0.549097842            & -1.877812726           & off  \\
44 & 233.4891  & 7.30065  & 0.10028 & n      & 3.623164361            & -0.009635092           & 0.001669191            & 0.56477776             & -2.099042247           & off  \\
45 & 233.75942 & 34.92733 & 0.13104 & m      & 1.757616661            & 0.136202228            & 0.028415326            & 0.499229127            & -0.842406462           & off  \\
46 & 234.73583 & 8.06631  & 0.12597 & n      & 3.367526298            & 0.073223346            & 0.007602986            & 0.520496031            & -2.064582136           & off  \\
47 & 236.37347 & 25.19107 & 0.1171  & n      & 3.269494139            & 0.013142188            & -0.001094589           & 0.577752239            & -1.922667145           & off  \\
48 & 238.09704 & 32.58194 & 0.12795 & m      & 2.814655555            & 0.061617556            & 0.000384853            & 0.514273055            & -1.77599174            & off  \\
49 & 238.24354 & 20.7872  & 0.13616 & n      & 3.684579878            & 0.00447027             & 0.008324203            & 0.60797451             & -2.095867425           & off  \\
50 & 242.41877 & 3.98624  & 0.06464 & n      & 2.863536339            & 0.029935537            & -0.001824506           & 0.510395418            & -1.816015226           & off  \\
51 & 244.77764 & 6.26993  & 0.14283 & n      & 3.453578941            & 0.031805787            & 0.013171976            & 0.580164265            & -2.0445176             & off  \\
52 & 245.01806 & 9.38436  & 0.10721 & n      & 3.047771882            & 0.00318396             & 0.025876111            & 0.5293959              & -1.88015634            & off  \\
53 & 247.27675 & 17.82834 & 0.14187 & n      & 3.201123117            & 0.035584388            & 0.02898894             & 0.542923658            & -1.931542921           & off  \\
54 & 247.99725 & 0.35659  & 0.07123 & n      & 2.680756193            & 0.025546582            & -0.000274124           & 0.493459906            & -1.746541127           & off  \\
55 & 248.01031 & 32.06094 & 0.05941 & n      & 3.123745583            & 0.218362166            & 0.040843581            & 0.561960367            & -1.883181568           & off  \\
56 & 248.36307 & 26.3181  & 0.13427 & n      & 3.012645743            & 0.003688588            & -0.005842863           & 0.519156025            & -1.846945957           & off  \\
57 & 249.55334 & 40.47452 & 0.07155 & n      & 3.790212923            & 0.027948635            & 0.022950603            & 0.586085012            & -2.124656781           & on   \\
58 & 250.0706  & 27.29643 & 0.11325 & n      & 2.872562685            & 0.031483693            & 0.028889777            & 0.497057032            & -1.857436721           & off  \\
59 & 253.62163 & 24.1279  & 0.1393  & m      & 3.586235895            & 0.157830551            & 0.042350039            & 0.584233174            & -1.973561102           & off  \\
60 & 253.64933 & 33.04761 & 0.12231 & n      & 3.836108848            & 0.127629575            & 0.000705294            & 0.575651169            & -2.16141279            & on   \\
61 & 332.6865  & 24.99946 & 0.12045 & n      & 3.225671231            & 0.007238851            & 0.001815825            & 0.53405058             & -1.992216163           & on   \\
62 & 340.30644 & -1.35248 & 0.05903 & m      & 4.138612087            & 0.02436495             & 0.010163016            & 0.607975117            & -2.303824046           & on   \\
63 & 342.58444 & 1.53832  & 0.1247  & n      & 3.554491203            & 0.147158822            & 0.009339877            & 0.542296857            & -2.071064578           & on     \\\hline
\end{tabular}
\tablecomments{(1) Object identifier (ID); (2) Right Ascension (RA, J2000); (3) Declination (DEC, J2000); (4) Redshift (z); (5) Merger classification (``m'' = merger, ``o'' = other); (6) Concentration ($C$); (7) Asymmetry ($A$); (8) Smoothness ($S$); (9) Gini coefficient ($G$); (10) $M_{20}$; (11) Morphological state defined in Section \ref{sec:sec3.4}.}
\end{table}

\section{Mergers Identified through Visual Classification }\label{sec:B1}

\begin{figure}[ht!]
    \centering
     \subfigure[]{
        \includegraphics[width=0.45\textwidth]{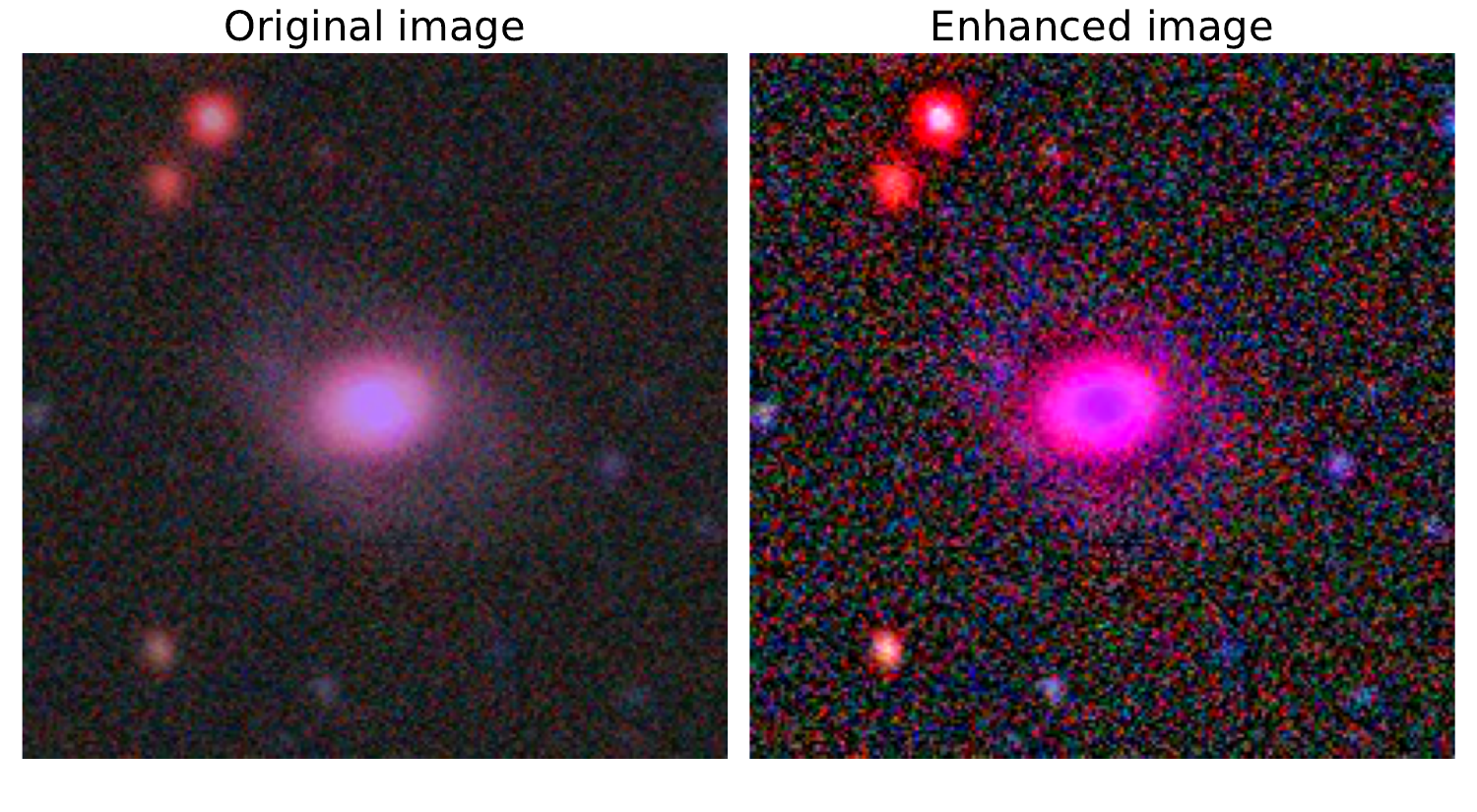}
    }
    \subfigure[]{
        \includegraphics[width=0.45\textwidth]{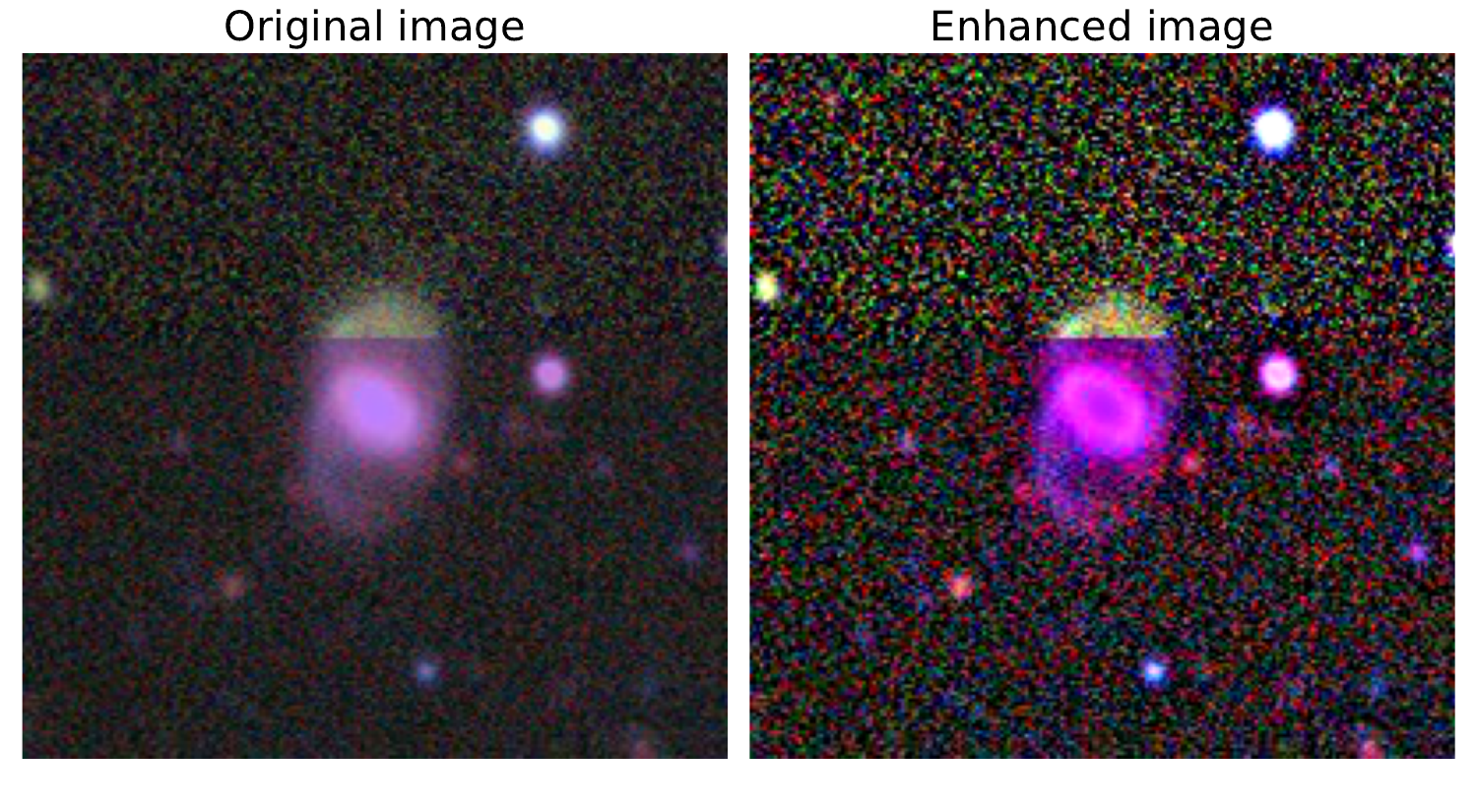}
    }
    \subfigure[]{
        \includegraphics[width=0.45\textwidth]{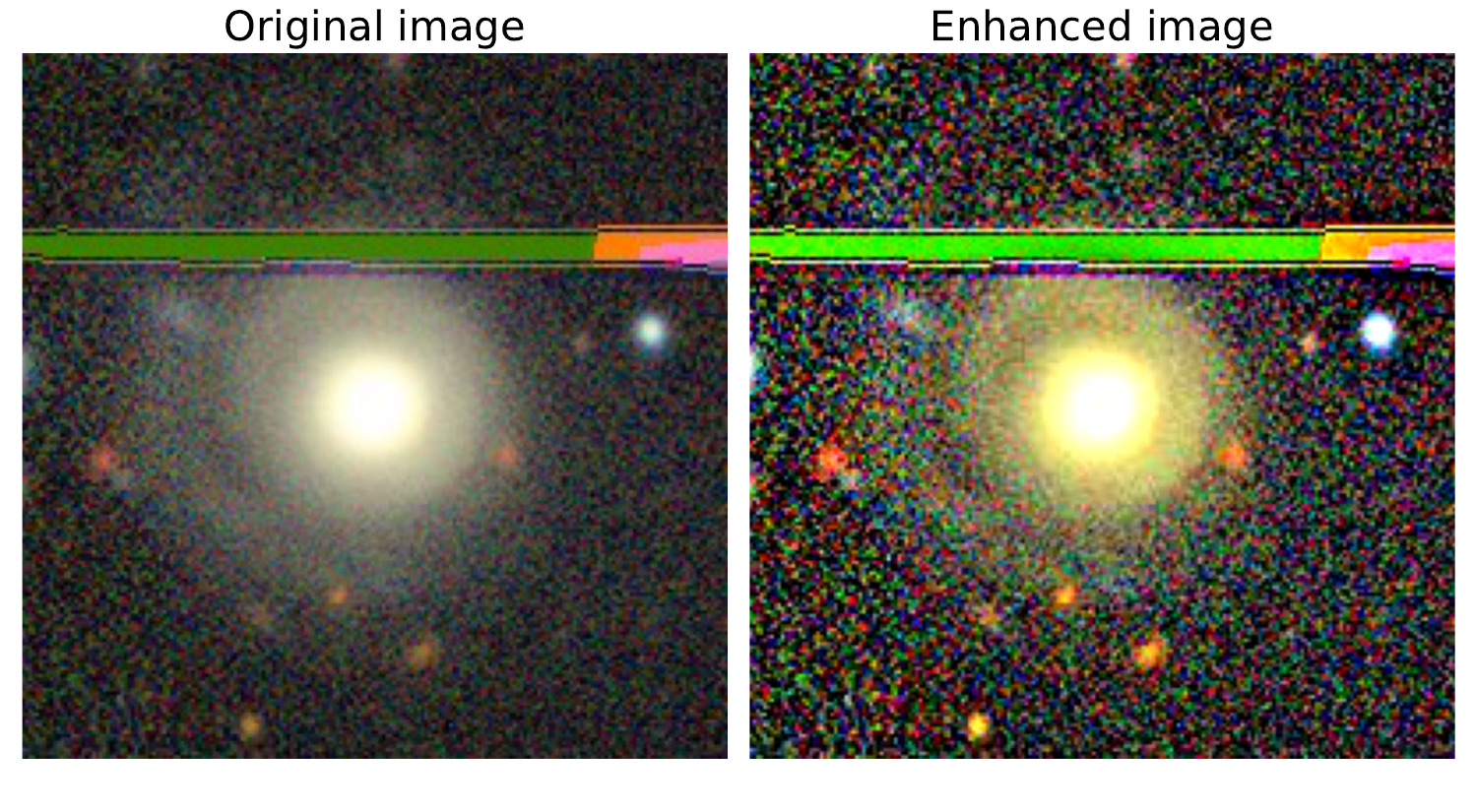}
    }
    \subfigure[]{
        \includegraphics[width=0.45\textwidth]{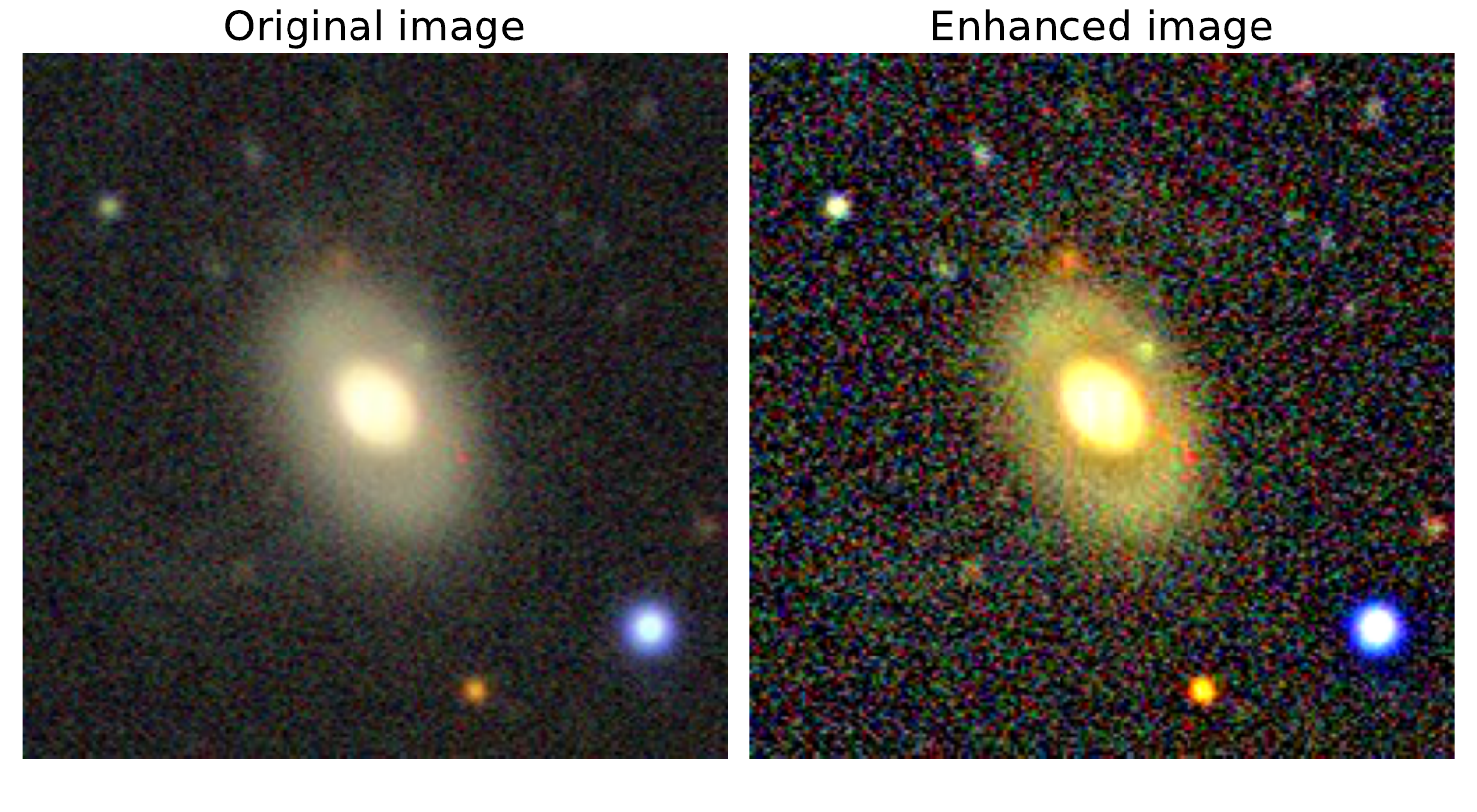}
    }
    \subfigure[]{
        \includegraphics[width=0.45\textwidth]{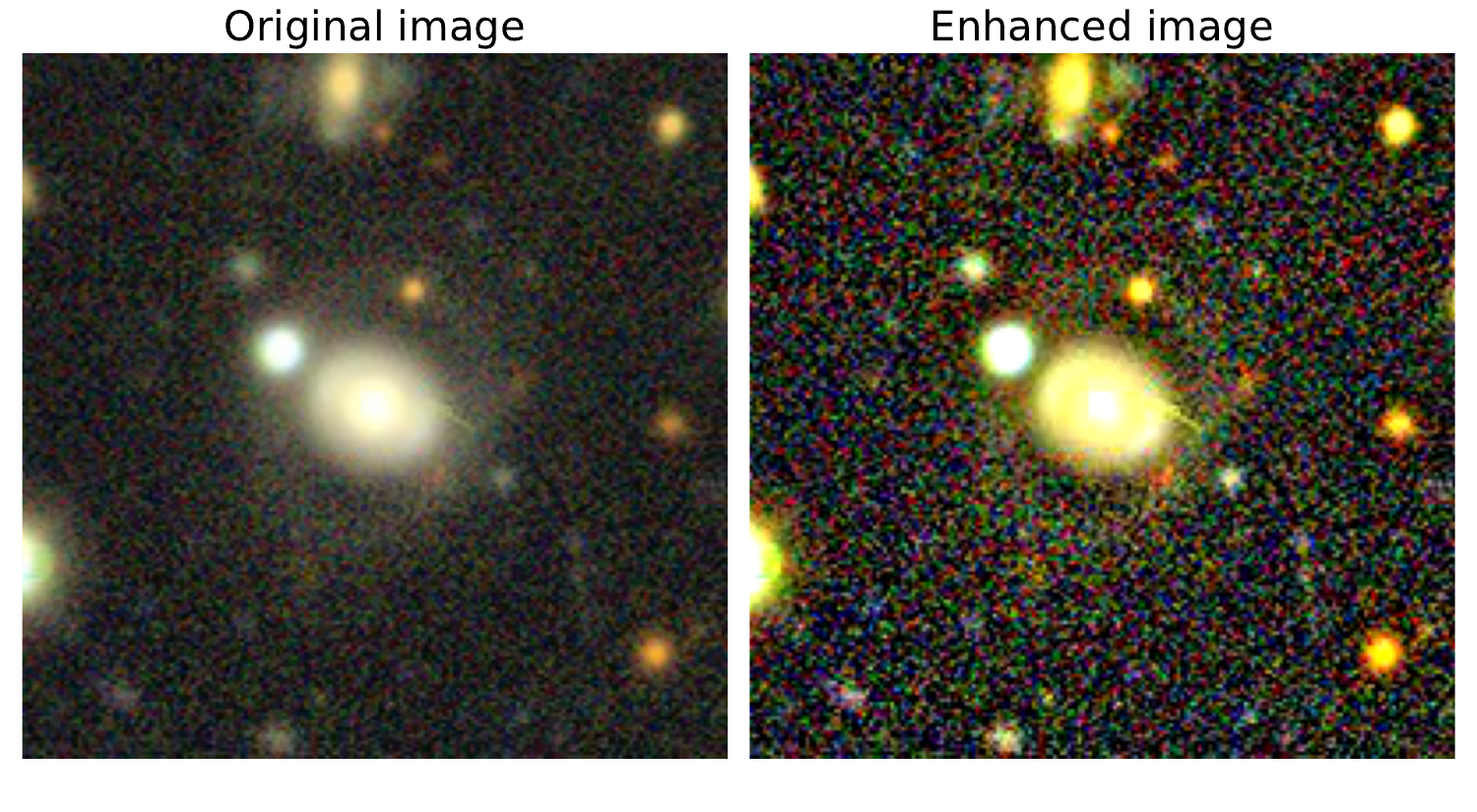}
    }
    \subfigure[]{
        \includegraphics[width=0.45\textwidth]{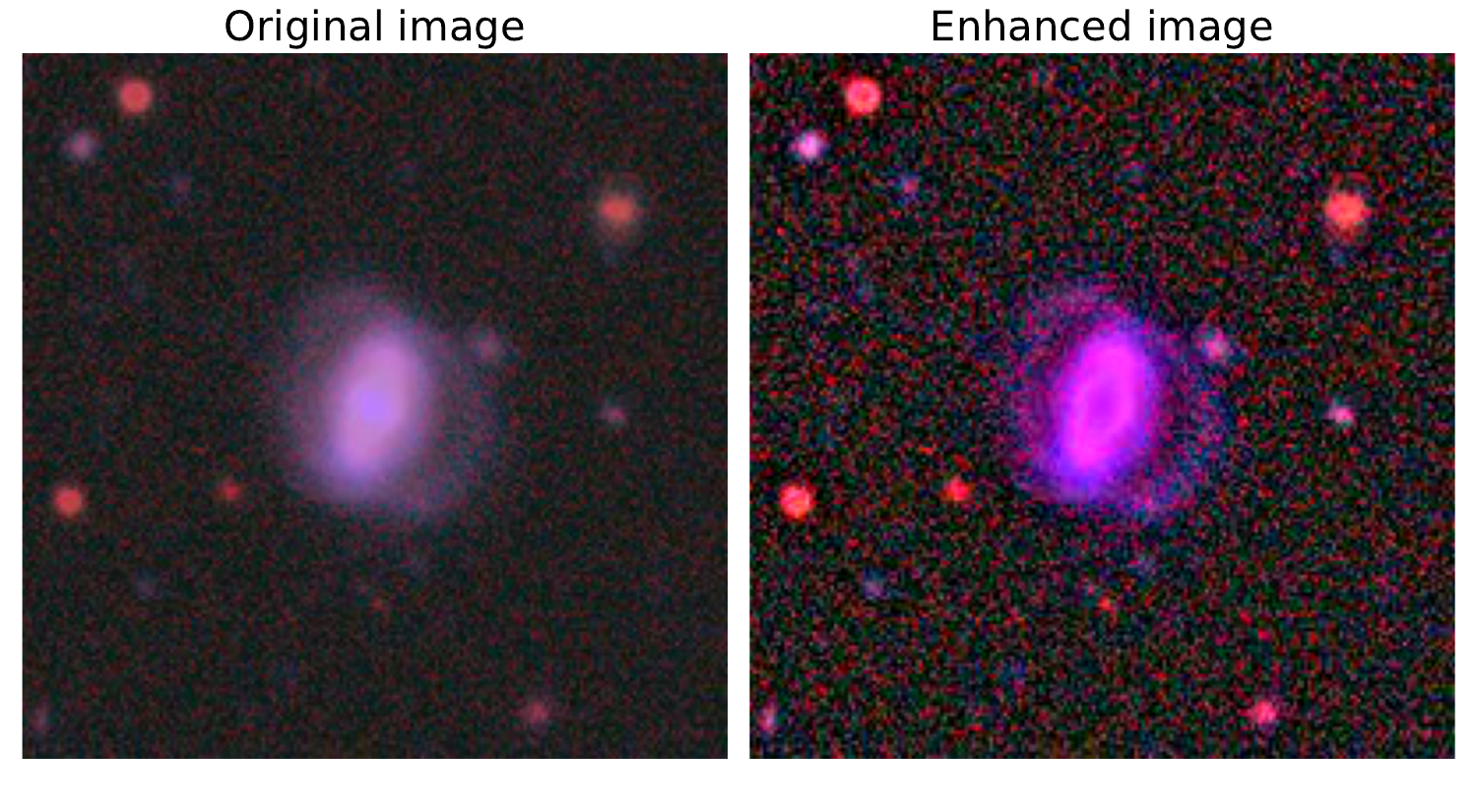}
    }
    \subfigure[]{
        \includegraphics[width=0.45\textwidth]{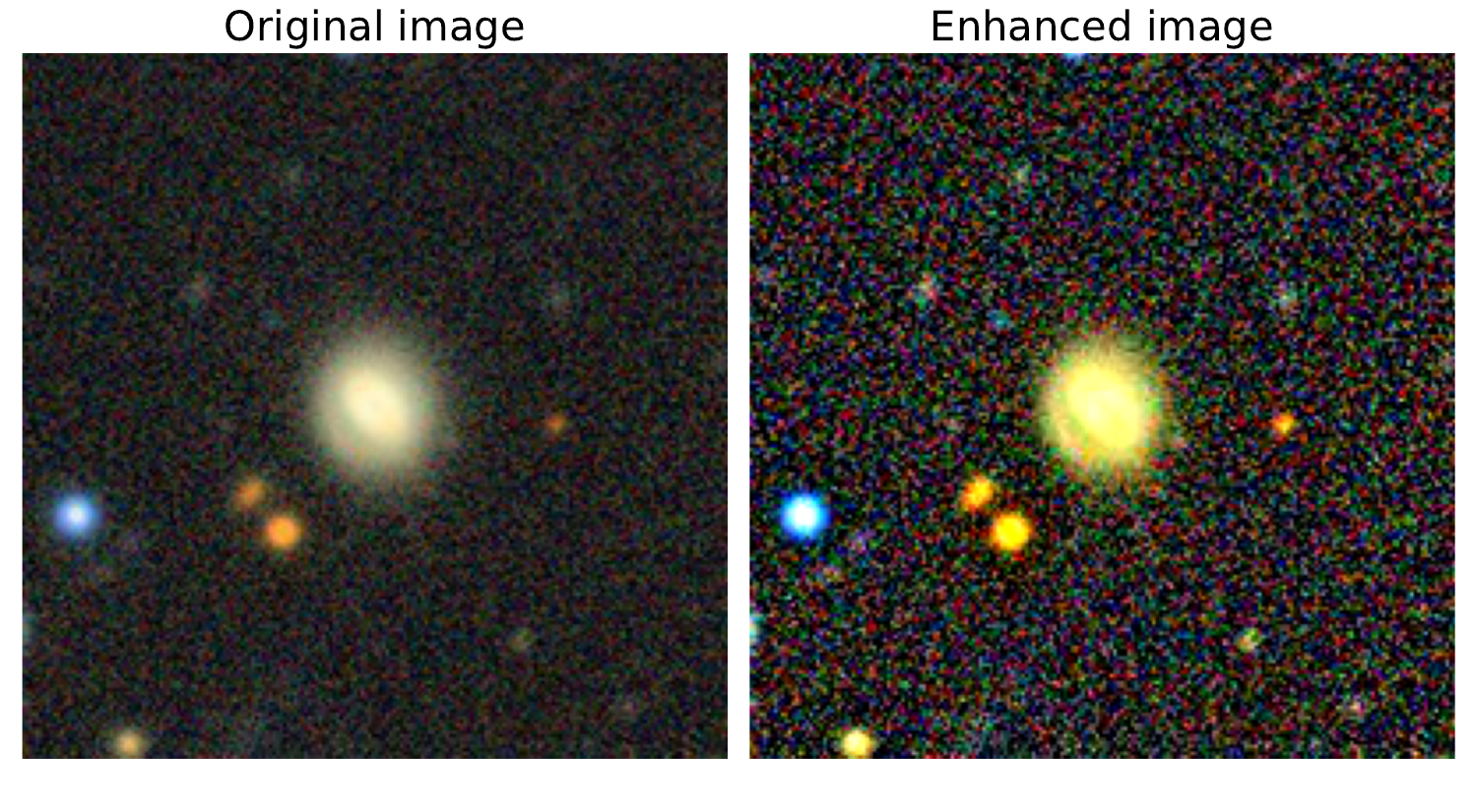}
    }
    \subfigure[]{
        \includegraphics[width=0.45\textwidth]{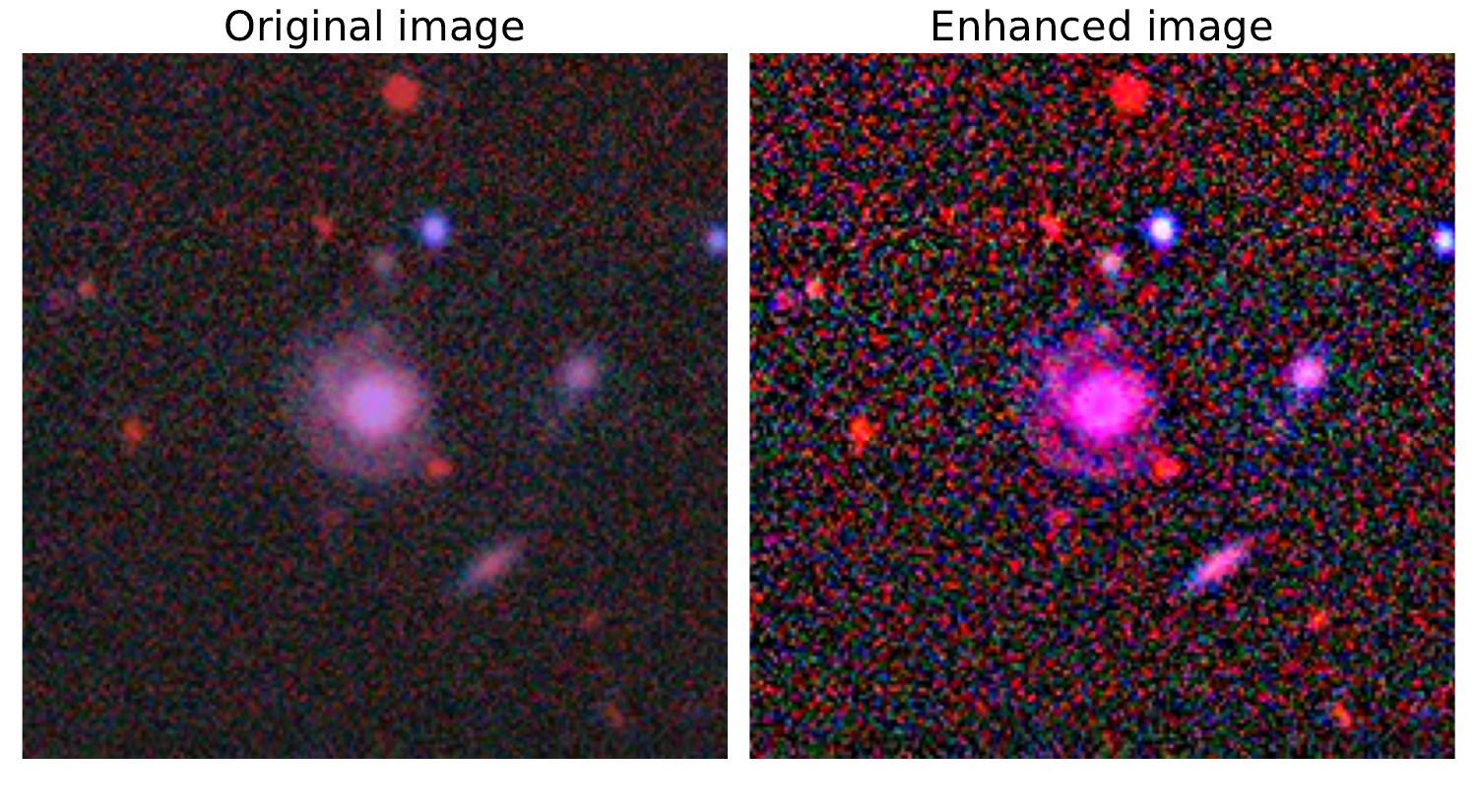}
    }
    
    \caption{Additional mergers of CL-AGN host galaxies via visual Classification.
    (a)-(i) show shell structures (ID 1,7,8,11,13,14,33,59,62); (j)-(m) show a visible large-scale asymmetry (ID 5,18,30,48); (n) displays a tidal feature (ID 2). (h), (i), and (k) are absent from our sample with available morphological parameters.}
\label{fig:fig4}
\end{figure}

\begin{figure}[t]
\figurenum{5}
    \centering
    \setcounter{subfigure}{8}
    \subfigure[]{
        \includegraphics[width=0.45\textwidth]{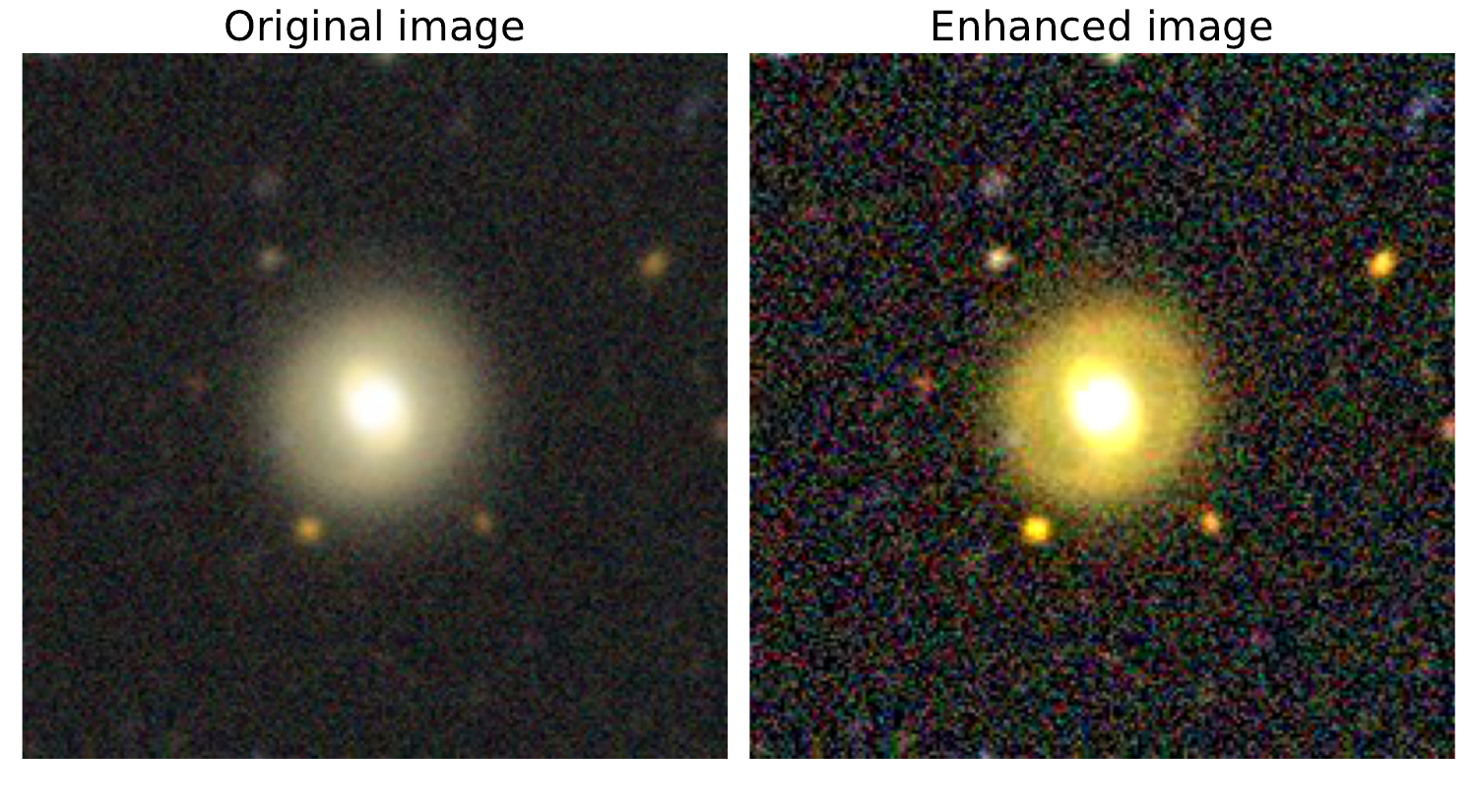}
    }
    \setcounter{subfigure}{9}
    \subfigure[]{
        \includegraphics[width=0.45\textwidth]{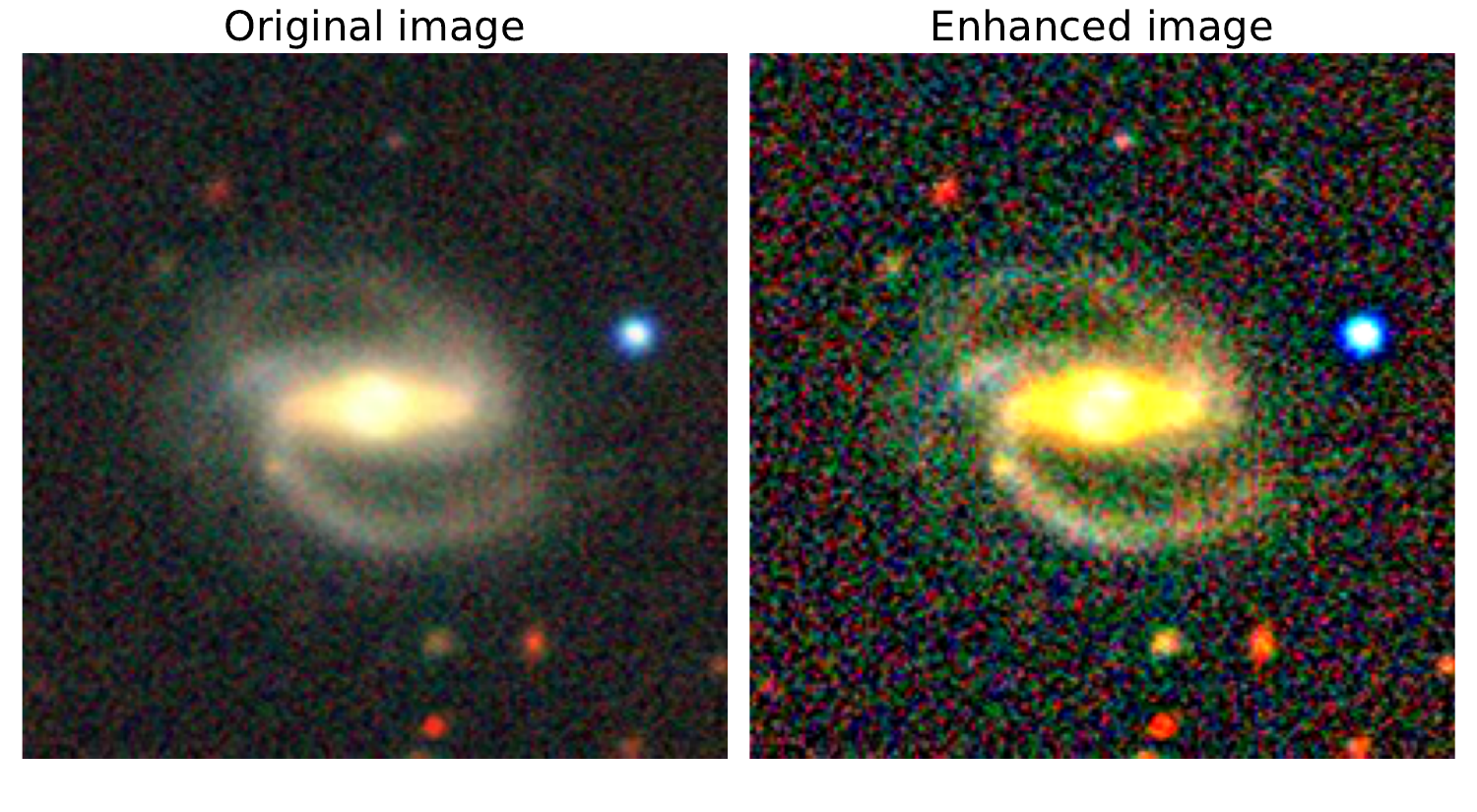}
    }
    \setcounter{subfigure}{10}
    \subfigure[]{
        \includegraphics[width=0.45\textwidth]{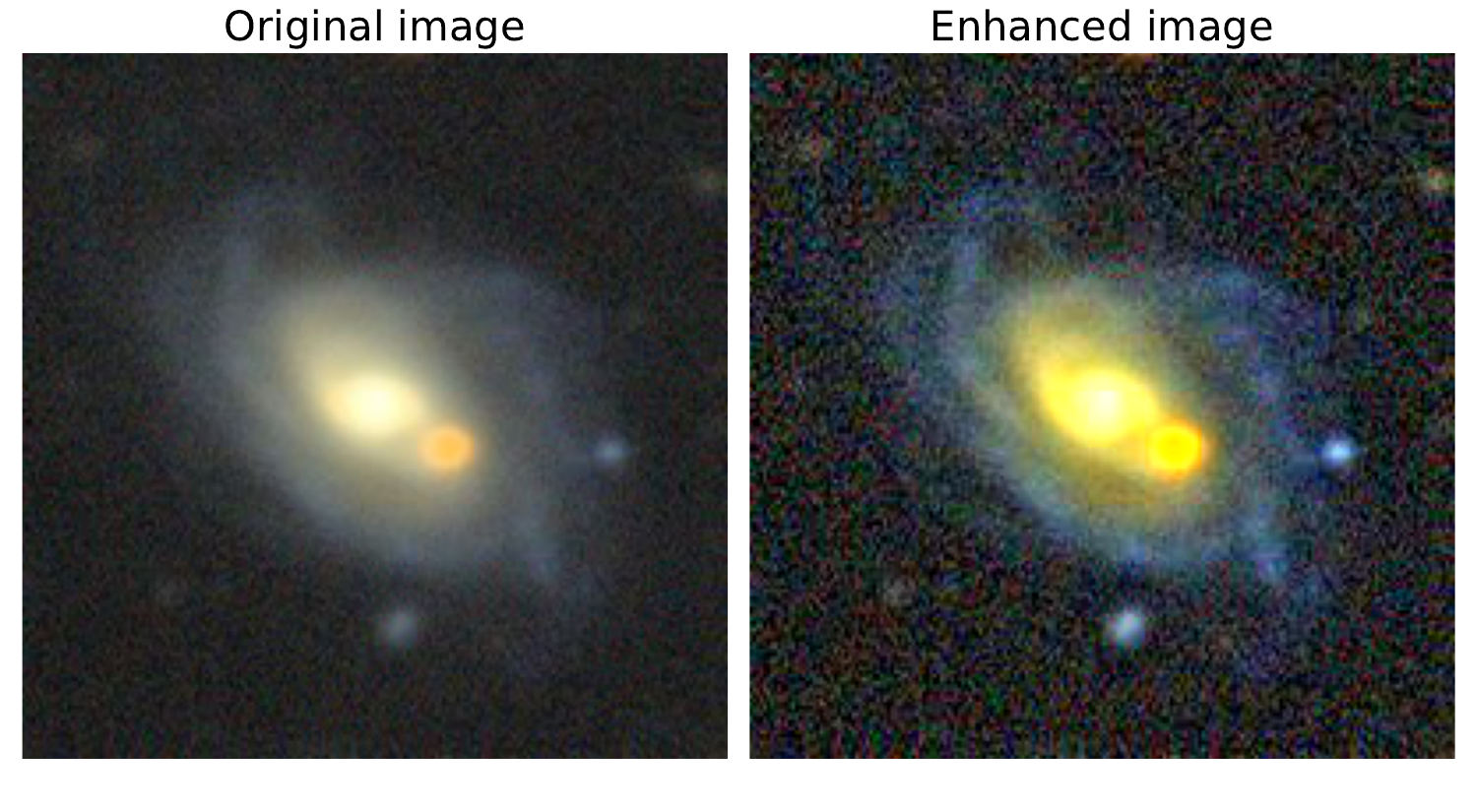}
    }
    \setcounter{subfigure}{11}
    \subfigure[]{
        \includegraphics[width=0.45\textwidth]{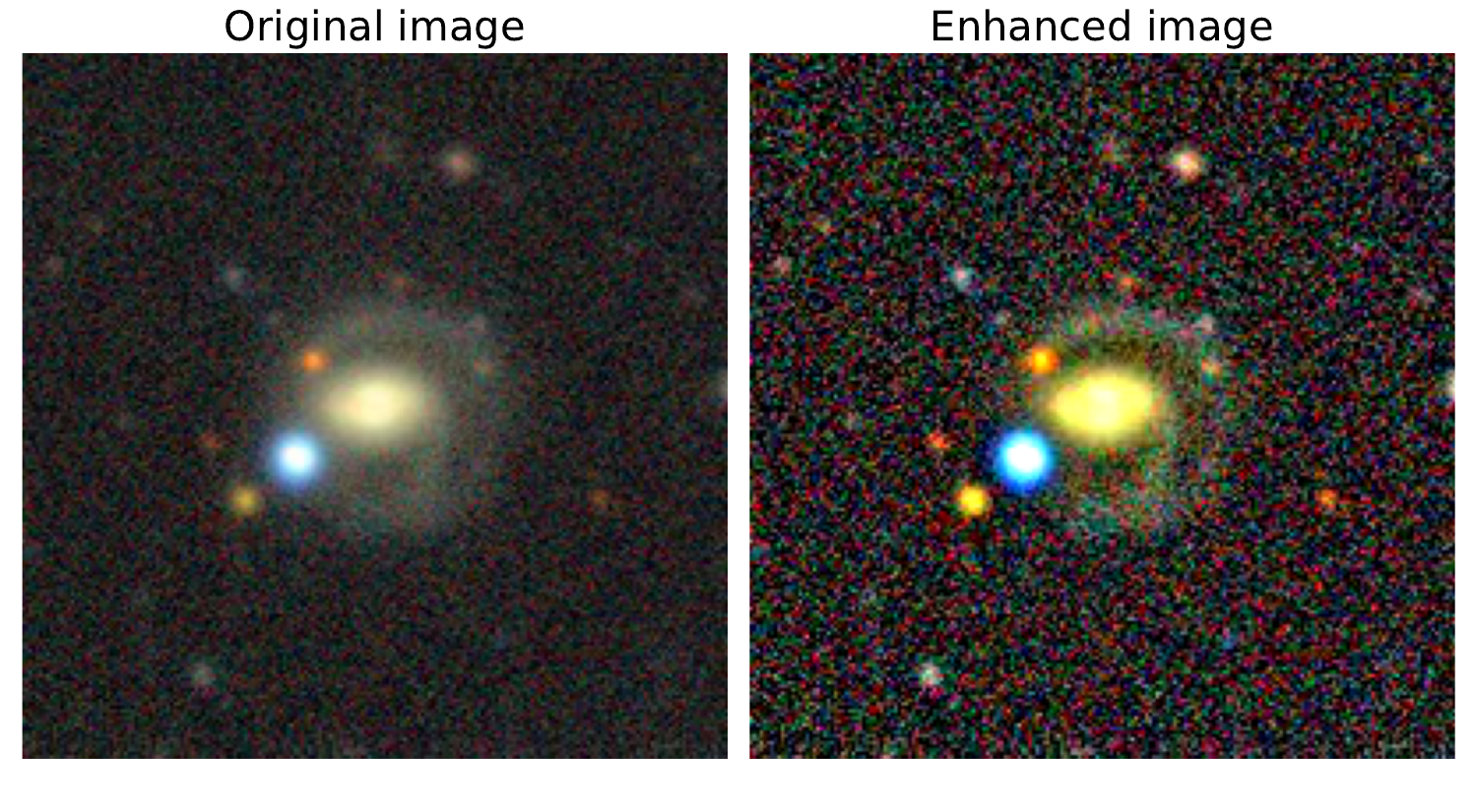}
    }
    \setcounter{subfigure}{12}
    \subfigure[]{
        \includegraphics[width=0.45\textwidth]{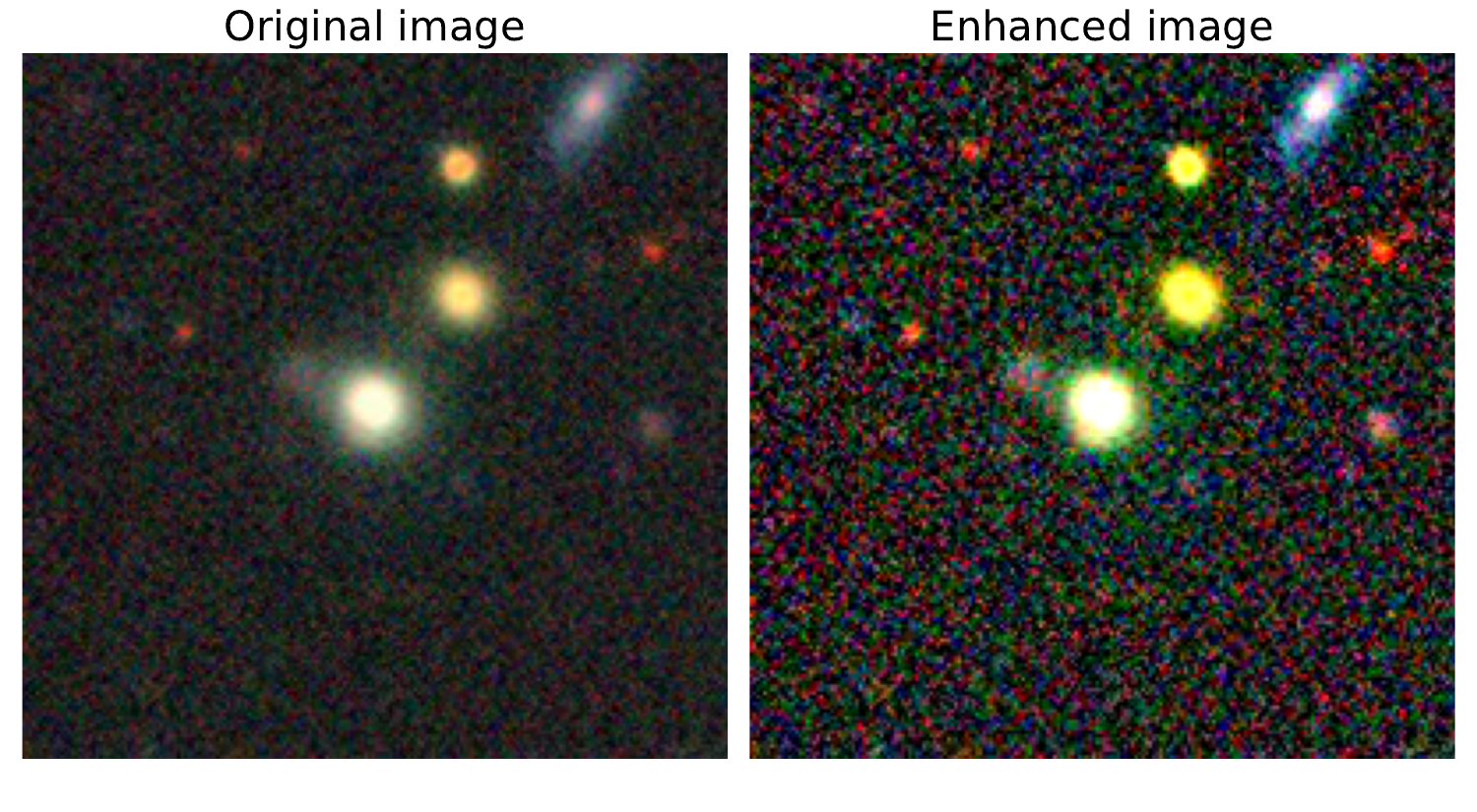}
    }
    \setcounter{subfigure}{13}
    \subfigure[]{
        \includegraphics[width=0.45\textwidth]{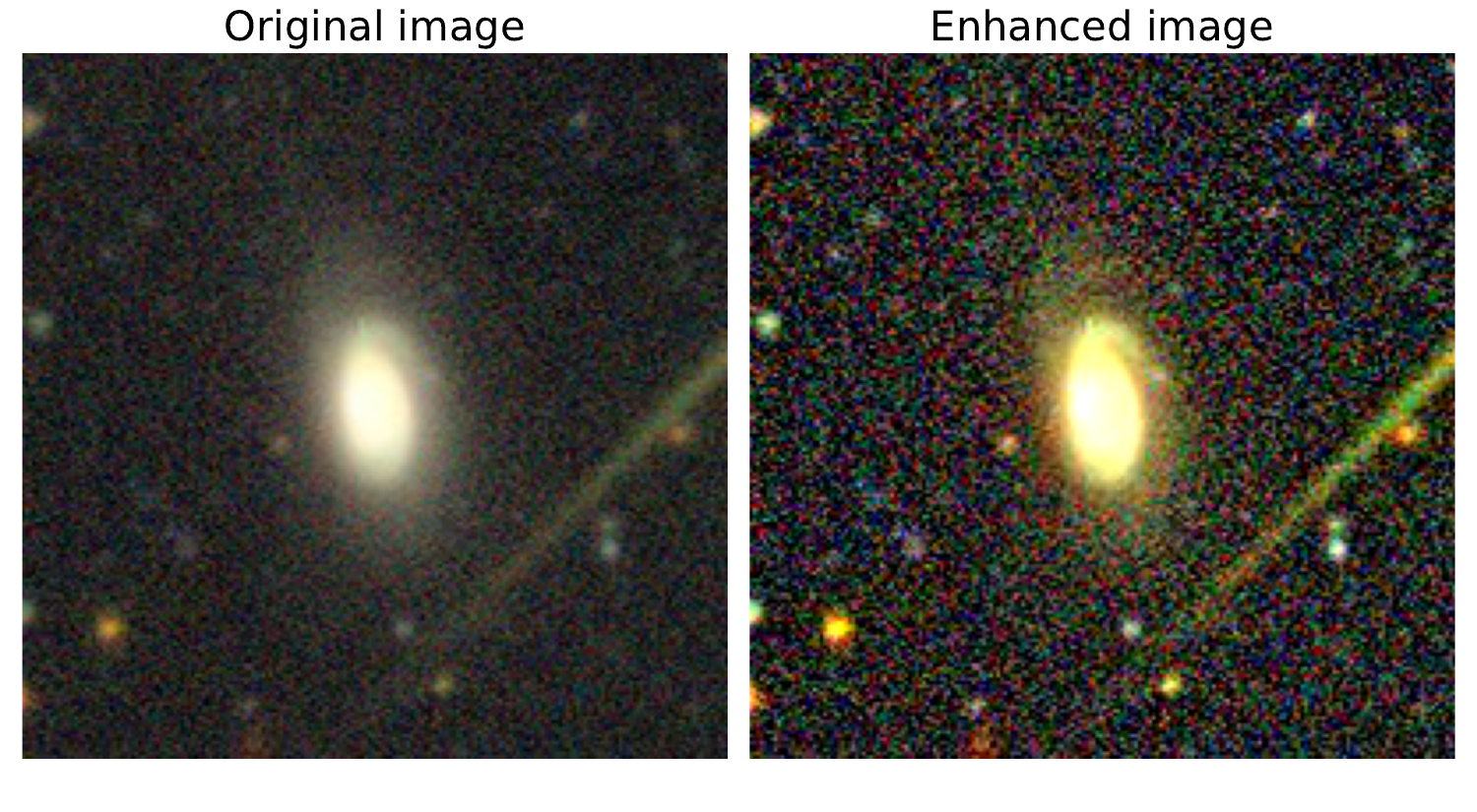}
    }
    \caption{(Continued.)}
\end{figure}

\bibliography{sample701}{}
\bibliographystyle{aasjournalv7}
\end{CJK*}
\end{document}